\documentclass[conference,compsoc]{IEEEtran}
\usepackage{amsmath}
\usepackage{amsthm}

\usepackage{cancel}
\usepackage{amssymb}
\usepackage{amsfonts}
\usepackage{utfsym} 
\usepackage{wasysym}
\usepackage[colorlinks=true,urlcolor=black]{hyperref}
\usepackage{cleveref}
\crefname{section}{Sec.}{Secs.}
\Crefname{section}{Section}{Sections}
\Crefname{table}{Table}{Tables}
\crefname{table}{Tab.}{Tabs.}
\Crefname{figure}{Figure}{Figures}
\crefname{figure}{Fig.}{Figs.}
\Crefname{equation}{Equation}{Equations}
\crefname{equation}{Eq.}{Eqs.}
\crefname{algorithm}{Alg.}{Algs.}
\Crefname{algorithm}{Algorithm}{Algorithms}
\usepackage{url} 
\usepackage{graphicx}
\usepackage{multirow}
\usepackage{makecell}
\usepackage[flushleft]{threeparttable}
\usepackage{colortbl}
\usepackage{booktabs}
\usepackage{xspace}
\usepackage[dvipsnames]{xcolor}
\usepackage{enumitem}
\usepackage[most]{tcolorbox} 
\usepackage{algorithm}
\usepackage{algorithmic}

\tcbset{
  takeaway/.style={
    colback=gray!10,       
    colframe=gray!60,      
    boxrule=0.8pt,         
    arc=4mm,               
    left=5mm, right=5mm,   
    top=3mm, bottom=3mm,   
    fonttitle=\bfseries,
    coltitle=black
  }
}

\usepackage{balance}

\graphicspath{{figures/}}

\newcommand{\tablefontsize}{\scriptsize}

\ifCLASSOPTIONcompsoc
  \usepackage[nocompress]{cite}
\else
  \usepackage{cite}
\fi
\ifCLASSINFOpdf
\else
\fi
\begin{document}
%
\title{Beyond Attack-Based Accuracy: Rethinking Adversarial Robustness Evaluation Through Structural Gradient Alignment}



%

\author{\IEEEauthorblockN{Hanrui Wang\IEEEauthorrefmark{1},
Yu Luo\IEEEauthorrefmark{2},
Shuo Wang\IEEEauthorrefmark{3}, 
Xingbo Dong\IEEEauthorrefmark{2},
Zhe Jin\IEEEauthorrefmark{2},
and Isao Echizen\IEEEauthorrefmark{1}}
\IEEEauthorblockA{\IEEEauthorrefmark{1}National Institute of Informatics, Tokyo, Japan\\email: hanrui\_wang@nii.ac.jp}
\IEEEauthorblockA{\IEEEauthorrefmark{2}Anhui University, Hefei, China~
}
\IEEEauthorblockA{\IEEEauthorrefmark{3}Shanghai Jiao Tong University, Shanghai, China~
}
}


\maketitle

\begin{abstract}
Visual adversarial robustness remains a fundamental security property as vision systems evolve from conventional classifiers to vision-language models (VLMs) and large vision-language models (LVLMs), with visual inputs remaining an attack surface across generations. Yet existing evaluation methods often depend on prescribed attacks, in-distribution data, extensive probing, or classifier-specific formulations, limiting external auditing of diverse models. We introduce \emph{Structural Gradient Alignment} (SGA), a complementary robustness metric measuring the structural alignment between input content and objective-conditioned local model sensitivity. SGA uses a few randomly sampled, unlabeled public probes that need not follow the downstream task distribution and only one input-gradient computation per model--probe pair. Across 38 distinct DNNs, VLMs, and LVLMs, SGA produces rankings consistent with attack-based references. Without robustness references, uncalibrated SGA achieves $83.72\%$ average pairwise agreement across eight public-probe conditions, outperforming the strongest overall baseline in all eight by $9.38$ percentage points on average. With known robustness orders of anchor models, calibrated SGA increases agreement to $85.36\%$. These results demonstrate an attack-agnostic, computationally efficient robustness signal across conventional DNNs, VLMs, and autoregressive LVLMs, downstream tasks, threat norms, probe distributions, and evaluator-knowledge settings.
\emph{Our implementation is available at \url{https://github.com/azrealwang/SGA}.}
\end{abstract}


%
\IEEEpeerreviewmaketitle

\section{Introduction}
\label{sec:introduction}

Adversarial manipulation of visual inputs remains a persistent attack surface across successive generations of vision systems. Modern vision-language models (VLMs) and large vision-language models (LVLMs) retain visual perception pathways susceptible to adversarial manipulation, extending the need for robustness evaluation beyond conventional image classifiers. Robustness evaluation must therefore evolve together with the models being evaluated. The dominant evaluation paradigm remains attack-based: evaluators instantiate a threat model, generate adversarial examples using attacks such as Projected Gradient Descent (PGD)~\cite{madry2018towards} or AutoAttack~\cite{croce2020reliable}, and report robust accuracy. Other studies estimate robustness through local Lipschitz properties~\cite{weng2018evaluating}, linearized approximations to decision-boundary distance~\cite{etmann2019connection}, generative models~\cite{li2024great}, foundation models~\cite{zhang2024foundation}, or architecture-level proxies~\cite{feng2025zero}. Despite their different mechanisms, existing methods have not fully kept pace with this model evolution: they generally retain substantial dependence on attack specifications, threat norms, in-distribution test sets, computational resources, or classifier-specific output structures.

Four gaps limit existing robustness evaluation in realistic model-auditing scenarios.

\textbf{Gap 1: Attack dependence.}
Robust accuracy depends on the selected attack, threat norm, and configuration~\cite{madry2018towards,croce2020reliable}, limiting evaluation when suitable attack protocols are unknown or difficult to instantiate. We therefore seek attack-agnostic evaluation through input sensitivity, without generating adversarial examples, executing prescribed attacks, or requiring target-model threat norms.

\textbf{Gap 2: Data dependence.}
Existing evaluators often require labeled, in-distribution test sets or extensive probing~\cite{goodfellow2014explaining,croce2020reliable,weng2018evaluating,etmann2019connection}. Such data may be private, restricted, or unavailable to external auditors, while extensive probing is impractical under limited data access. We instead aim to evaluate models using few randomly sampled, unlabeled public probes independent of downstream task distributions.

\textbf{Gap 3: Computational cost.}
Iterative attacks and sampling-based estimators require repeated forward and backward passes~\cite{croce2020reliable,weng2018evaluating}, making evaluation costly for large model collections or computationally intensive models such as LVLMs. We therefore avoid iterative optimization and extensive gradient sampling to enable efficient evaluation.

\textbf{Gap 4: Evaluation has not kept pace with model evolution.}
Many evaluators rely on classifier-specific outputs, class margins, or decision boundaries~\cite{weng2018evaluating,etmann2019connection}, which do not naturally extend to VLMs and autoregressive LVLMs with different output interfaces and objectives. We seek a consistent robustness-evaluation framework across conventional classifiers, VLMs, and autoregressive LVLMs.

Together, these objectives define a new robustness-auditing setting that reduces dependence on prescribed attacks and in-distribution test sets while remaining applicable across evolving model interfaces. This setting could support third-party model auditing under limited information disclosure, preliminary screening of competition submissions without exposing private test sets, and large-scale checkpoint triage before comprehensive testing. We position it as a scalable first stage of robustness evaluation that can be supplemented, where appropriate, by comprehensive attack-based evaluation or formal certification.

\begin{figure}[!t]
    \centering
    \includegraphics[width=\linewidth]{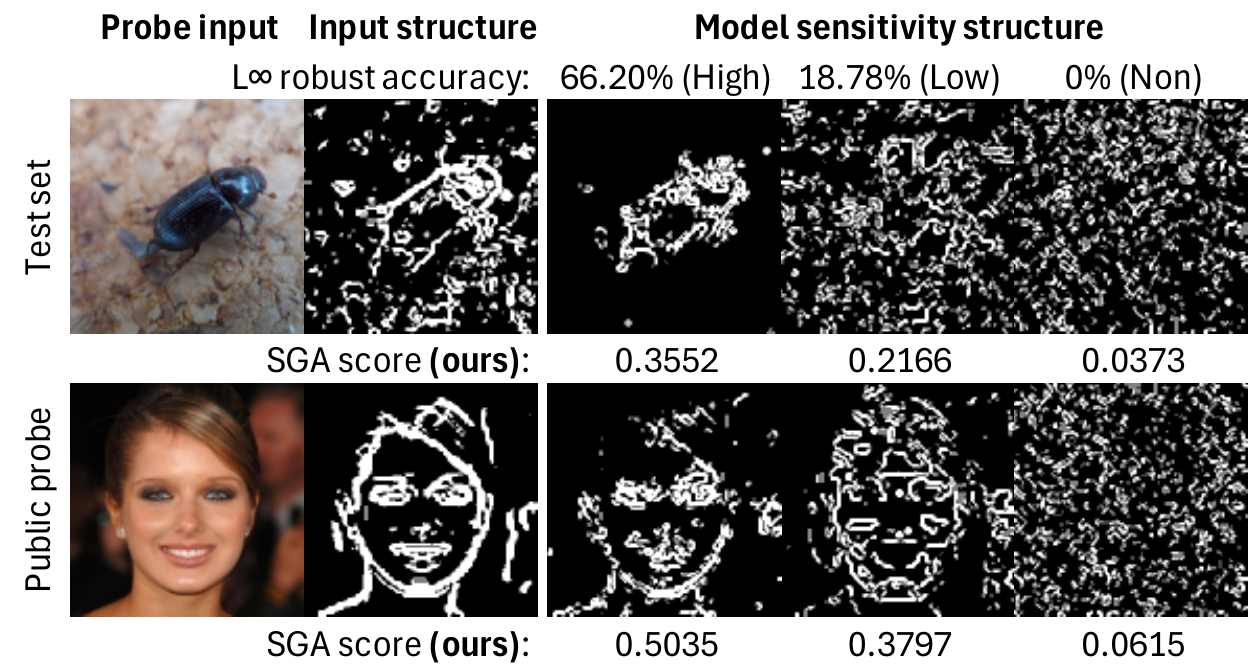}
    \caption{\textbf{Illustration of SGA across robustness levels and probe sources.} SGA measures structural alignment between input structure and objective-conditioned local sensitivity. Across test-set and task-independent public probes, higher-reference-robustness models exhibit stronger alignment and higher SGA scores. Reference $\ell_\infty$ PGD robust accuracies are shown only for comparison and are not used by SGA.}
    \label{fig:intro}
\end{figure}

To instantiate this setting, we propose \emph{Structural Gradient Alignment} (SGA), motivated by a \textbf{structural sensitivity hypothesis}: adversarially robust models tend to organize local input sensitivity around stable spatial structure, whereas weaker robustness is more often associated with spatially incidental or fragmented sensitivity. Prior work links robustness to interpretable input gradients~\cite{etmann2019connection}, but its task-specific formulation does not naturally support task-independent probes or diverse model interfaces. SGA formalizes the input gradient as an objective-conditioned local sensitivity field, maps both the input and this field into a common spatial-structure domain, and measures their alignment within that domain, as illustrated in \cref{fig:intro}. This formulation compares where salient local variation occurs rather than directly comparing heterogeneous image and gradient values. A spatial-structure destruction experiment further supports this hypothesis, showing that disrupting gradient spatial organization substantially reduces robustness-ranking agreement while preserving non-spatial gradient statistics.

SGA supports three protocols for different auditing conditions. A model pool contains models with directly comparable reference robustness under the same task, threat model, and attack protocol. \emph{Transfer SGA} applies a preset selected on a nonoverlapping development model set, providing stable, comparable scores without target-pool references; \emph{uncalibrated SGA} selects a configuration from the evaluated models alone to improve within-cohort ranking; and \emph{calibrated SGA} uses the known robustness order of same-pool anchors to align subsequent evaluations with an established ranking standard. None accesses the target models' reference robustness.

We evaluate three SGA protocols on 38 distinct systems across four homogeneous pools (DNN $\ell_2$, DNN $\ell_\infty$, VLM $\ell_\infty$, and LVLM $\ell_\infty$) and four probe datasets: two in-distribution test sets and two task-independent public datasets. We assess agreement with PGD- and AutoAttack-based robustness rankings. Across eight public-probe conditions, Transfer SGA achieves $83.48\%$ average pairwise agreement (PA), exceeding the strongest overall baseline by $9.15$ percentage points. Uncalibrated SGA selects configurations without robustness references, raising PA to $83.72\%$, while Calibrated SGA uses known anchor-model orders to reach $85.36\%$. Transfer SGA ranks five LVLMs with ten public probes in $103.8$ seconds on one H100 GPU; uncalibrated and calibrated selection add at most $0.032$ seconds by reusing cached gradients for structural post-processing. Among evaluated methods~\cite{goodfellow2014explaining,weng2018evaluating,etmann2019connection,li2024great}, only SGA consistently ranks classifiers and LVLMs with diverse visual encoders under a shared language interface using task-independent public probes.

We make three contributions:
\begin{itemize}[leftmargin=*]
    \item \textbf{A realistic robustness-auditing setting.}
    We evaluate target models using input gradients and few unlabeled, task-independent public probes.

    \item \textbf{A structural robustness metric.}
    SGA measures alignment between input content and objective-conditioned local sensitivity in a common spatial-structure domain with one gradient per model--probe pair. We characterize finite-probe concentration and sufficient conditions for preserving SGA order across probe distributions.

    \item \textbf{Three evaluation protocols.}
    Transfer SGA uses a transferred preset for comparable scoring; uncalibrated SGA improves ranking without robustness references; calibrated SGA uses ordered anchors to align with an established ranking standard. None requires target-model reference robustness, and both adaptive protocols use lightweight post-processing of cached gradients.
\end{itemize}

\noindent\textbf{Our position.}
Adversarial robustness should not be equated with accuracy under a particular attack protocol. Conventional attack-based evaluation measures robustness within predefined threat models, often specified by $\ell_p$ norms and fixed perturbation budgets. Yet deployment conditions and future attacks cannot always be anticipated, limiting what such measurements reveal about robustness beyond their assumptions. We therefore argue for complementary evaluation of model properties without prescribing attack conditions. SGA takes this perspective by measuring structural alignment between visual content and objective-conditioned input sensitivity, without requiring attack algorithms, perturbation norms, or budgets. This measurement is distinct from attack-based robust accuracy, and its consistency with established references is evaluated empirically rather than assumed. Establishing a general theoretical connection between structural alignment and adversarial robustness remains an open challenge.

\section{Related Work}
\label{sec:related}

\cref{tab:related} compares representative robustness-evaluation methods by attack dependence, probe requirements, LVLM support, and computational cost. FGSM, PGD, and AutoAttack evaluate robustness by generating adversarial examples under specified threat models~\cite{goodfellow2014explaining,madry2018towards,croce2020reliable}, while RobustBench standardizes such evaluation~\cite{croce2021robustbench}. These methods provide protocol-specific robust accuracy but require attack specification, test data, and adversarial-example generation. SGA provides a complementary attack-agnostic signal; we use PGD- and AutoAttack-induced orders only as protocol-dependent references. CLEVER and Linearized Robustness estimate norm-specific or linearized decision-boundary distances from classifier margins and gradients~\cite{weng2018evaluating,etmann2019connection}. GREAT and FMR require auxiliary generative or foundation models~\cite{li2024great,zhang2024foundation}; we exclude FMR because no public implementation is available. Zero-cost robust NAS~\cite{feng2025zero}, gradient-alignment analyses~\cite{ganz2023perceptually,rodriguez2024characterizing}, and AdvSMs~\cite{wang2026shared} address architecture selection, training, or sensitivity analysis rather than post-hoc model-level robustness evaluation. SGA instead supports few-shot evaluation with task-independent probes across classifiers and LVLMs without norm-specific output geometry or auxiliary models. Appendix~\ref{app:method-comparison} details these differences and LVLM adaptations.

\begin{table}[!t]
\centering
\begin{threeparttable}
\caption{\textbf{Comparison of robustness-evaluation methods.}}
\label{tab:related}
\tablefontsize
\setlength{\tabcolsep}{2.6mm}
\begin{tabular}{lccccc}
\toprule
\textbf{Method}
& \shortstack{\textbf{Attack-}\\\textbf{agnostic}}
& \shortstack{\textbf{Few-}\\\textbf{shot}}
& \shortstack{\textbf{Public}\\\textbf{probes}}
& \shortstack{\textbf{LVLM}\\\textbf{support}}\tnote{b}
& \textbf{Runtime}\tnote{c} \\
\midrule
FGSM~\cite{goodfellow2014explaining}
& $\usym{2717}$
& $\usym{2717}$
& $\usym{2717}$
& $\usym{2713}$
& 19.4\,s \\

AutoAttack~\cite{croce2020reliable}
& $\usym{2717}$
& $\usym{2717}$
& $\usym{2717}$
& $\usym{2713}$
& 5.4\,h \\

GREAT~\cite{li2024great}
& $\usym{2713}$
& $\usym{2717}$
& $\usym{2717}$
& $\usym{2717}$
& 16.3\,s \\

CLEVER~\cite{weng2018evaluating}
& $\usym{2713}$
& $\LEFTcircle$\tnote{a}
& $\LEFTcircle$
& $\usym{2717}$
& 4\,min \\

Linearized~\cite{etmann2019connection}
& $\usym{2713}$
& $\LEFTcircle$
& $\LEFTcircle$
& $\usym{2717}$
& 27.0\,s \\
\midrule
SGA (ours)
& \textbf{$\usym{2713}$}
& \textbf{$\usym{2713}$}
& \textbf{$\usym{2713}$}
& \textbf{$\usym{2713}$}
& 18.2\,s \\
\bottomrule
\end{tabular}

\begin{tablenotes}[flushleft]
\scriptsize
\item[a] $\LEFTcircle$ indicates a capability not claimed or validated in the original paper but effective after adaptation to our setting.
\item[b] LVLM support denotes reliable evaluation after adaptation, not mere executability.
\item[c] Measured on one H100 GPU. AutoAttack uses 1,000 inputs for one \texttt{Amini2024MeanSparse\_Swin-L}~\cite{amini2024meansparse} model; other methods use ten public probes per model to rank all 14 DNN $\ell_\infty$ models. Uncalibrated and calibrated SGA each add 0.021\,s after gradient caching.
\end{tablenotes}
\end{threeparttable}
\end{table}

\section{Preliminaries}
\label{sec:preliminaries}

\subsection{Threat Model and Evaluator Knowledge}
\label{sec:threat_model}

We study post-hoc robustness auditing with limited evaluator knowledge.
\begin{itemize}[leftmargin=*]
    \item \textbf{Objective.} Given models, the evaluator assigns continuous scores that may induce within-pool rankings.
    \item \textbf{Capabilities.} The evaluator can query outputs and compute input gradients but neither modifies models nor generates adversarial examples.
    \item \textbf{Information constraints.} No labeled test sets, training or threat-model metadata, or target-model reference robustness are required; only a few unlabeled, task-independent public probes and protocol-specific references are used.
    \item \textbf{Adversarial scope.} The evaluator randomly selects public probes, and SGA uses their gradients without generating adversarial perturbations. Attackers control neither probes nor scoring, so influencing SGA scores requires modifying the evaluated model's structural sensitivity. Prior work links gradient alignment to adversarial robustness~\cite{ganz2023perceptually,rodriguez2024characterizing}, suggesting that increasing structural alignment may also improve the property SGA aims to assess.
\end{itemize}

As summarized in \cref{tab:sga_protocols}, SGA supports three deployable protocols with different configuration-selection knowledge and a diagnostic \emph{oracle} using the complete target-model robustness ranking as an empirical upper bound.

\begin{table}[!t]
    \centering
    \tablefontsize
    \begin{threeparttable}
        \caption{\textbf{SGA protocols under different knowledge settings.}}
        \label{tab:sga_protocols}
        \setlength{\tabcolsep}{3.9mm}
        \begin{tabular}{lll}
            \toprule
            \textbf{Protocol}
            & \textbf{Knowledge}
            & \textbf{Criterion}\\
            \midrule
            Uncalibrated
            & None
            & Score dispersion \\

            Transfer
            & Development ranking
            & Development PA \\

            Calibrated
            & Ordered anchors
            & Anchor PA, then dispersion \\
            \midrule
            Oracle
            & Target ranking\tnote{a}
            & Target PA \\
            \bottomrule
        \end{tabular}

        \begin{tablenotes}[flushleft]
            \scriptsize
            \item[a] The oracle is an empirical upper bound over the evaluated batches and configurations, not a deployable protocol.
        \end{tablenotes}
    \end{threeparttable}
\end{table}

\subsection{Model Pools and Reference Evaluation}
\label{sec:model_pools}

Let $\mathcal{M}=\{f_1,\ldots,f_M\}$ denote a homogeneous model pool, defined operationally as a set of models whose robustness can be fairly ranked under the same attack-based reference evaluation. This evaluation fixes downstream task, perturbation norm and budget, attack configuration, evaluation inputs, scoring rule, and compatible interfaces. Homogeneity refers only to this common basis for reference evaluation and assumes no shared architecture, training procedure, or training-time threat model or perturbation budget. We restrict comparisons to homogeneous model pools and make no claims about cross-pool comparisons, including across perturbation norms (e.g., $\ell_2$ versus $\ell_\infty$) or downstream tasks (e.g., visual question answering (VQA)~\cite{antol2015vqa,lin2014microsoft} versus ImageNet classification~\cite{nips2017}).

Let $R_i$ denote the robustness score assigned to model $f_i$ by a fixed reference evaluation $\mathcal{A}$. This evaluation induces the reference relation
\begin{equation}
    f_i \succeq_{\mathcal{A}} f_j
    \quad\Longleftrightarrow\quad
    R_i \geq R_j.
    \label{eq:reference_order}
\end{equation}

We instantiate $\mathcal{A}$ using 50-step PGD (PGD-50) under each pool's task and threat-model settings or standard AutoAttack. For PGD-50, we retain only labeled test samples correctly handled by every model, yielding $100\%$ clean accuracy on the retained subset, and then attack those same samples. Thus, $R_i$ is either PGD-50 robust accuracy or the standard AutoAttack robust accuracy reported by RobustBench~\cite{croce2021robustbench}. We deliberately refer to the relation in \cref{eq:reference_order} as a \emph{reference order} for computing agreement. PGD-50 and AutoAttack are strong and widely adopted empirical evaluators, but we do not treat either as a universal ground truth for robustness.

\subsection{Few-shot Probe Settings}
\label{sec:probes}

The evaluator receives a visual probe batch
\begin{equation}
    \mathcal{D}_K=\{x_n\}_{n=1}^{K}.
\end{equation}
For LVLM evaluation, each probe image $x_n$ is additionally paired with a question prompt $q_n$ from the VQA dataset. We consider two probe settings:
\begin{itemize}
    \item \textbf{Test-set probes} are drawn from the labeled test set and provide ground-truth targets.
    \item \textbf{Public probes} are unlabeled, task-independent images that need not follow the downstream task distribution.
\end{itemize}
Within each pool and probe setting, every model and evaluation method receives the same probe images and, for LVLM evaluation, the same VQA question prompts.

\subsection{Evaluation Metric}
\label{sec:evaluation_metric}

SGA assigns each model $f_i$ a score $S_i$, where a larger value indicates stronger structural sensitivity alignment. We measure consistency with the reference order using pairwise agreement (PA). Let $\Delta S_{ij}=S_i-S_j$ and $\Delta R_{ij}=R_i-R_j$. For each model pair, define
\begin{equation}
\begin{aligned}
    a_{ij}
    &=
    \begin{cases}
        1, & \Delta R_{ij}=0,\\
        1, & \Delta S_{ij}\Delta R_{ij}>0,\\
        0, & \text{otherwise},
    \end{cases}
    \\[2pt]
    \mathrm{PA}(\mathbf{S},\mathbf{R})
    &=
    \frac{1}{\binom{M}{2}}
    \sum_{1\leq i<j\leq M}a_{ij}.
\end{aligned}
\label{eq:pa}
\end{equation}

Reference ties are counted as agreements because the reference evaluation does not distinguish between the tied models. However, the reference scores are distinct within every evaluated pool, so this convention does not affect our results. Ties in SGA scores are counted as disagreements. PA lies in $[0,1]$, is computed separately within each homogeneous model pool, and quantifies agreement with the reference order induced by the selected reference evaluation rather than with a universal ground truth for robustness.

\section{Structural Gradient Alignment (SGA)}
\label{sec:method}

SGA evaluates robustness through the structural correspondence between an input image and the model sensitivity induced by that input. Given a model $f_i$ and a probe batch $\mathcal{D}_K$, SGA constructs an evaluation target for each probe, computes the corresponding input gradient, and transforms the image and gradient into structural maps for comparison. SGA does not generate adversarial examples and requires only one input-gradient computation per model--probe pair.

\begin{figure*}[!t]
    \centering
    \includegraphics[width=\linewidth]{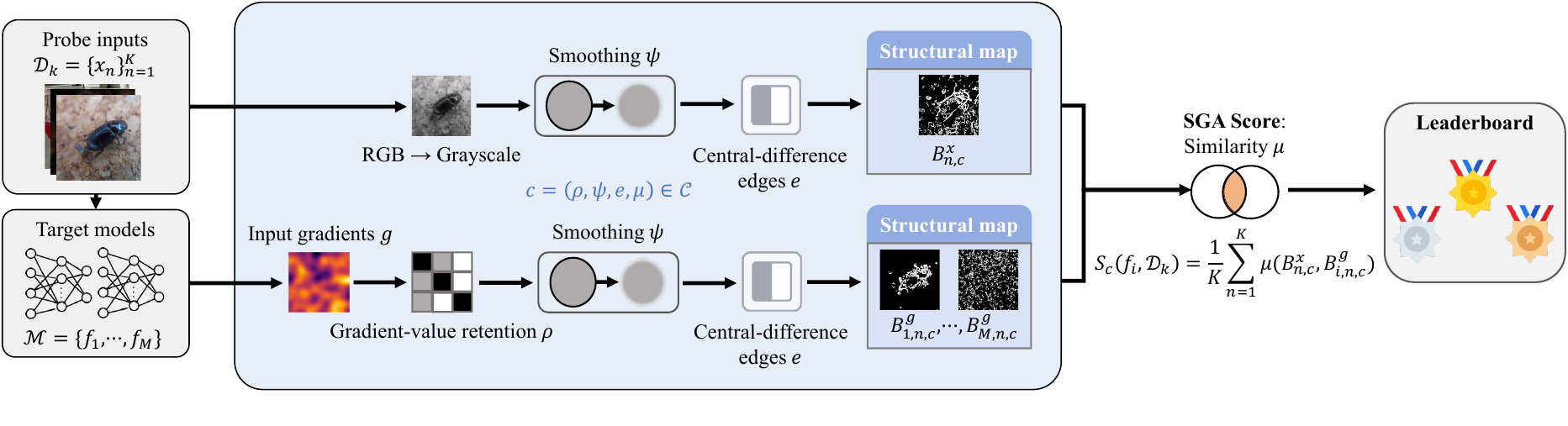}
    \caption{\textbf{SGA scoring pipeline.} For each model--probe pair, SGA transforms the probe image and its input gradient into structural maps under configuration $c=(\rho,\psi,e,\mu)$. It then measures their structural correspondence and aggregates the resulting scores across the probe batch to rank the evaluated models.}
    \label{fig:method}
\end{figure*}

Structural post-processing follows a configuration
\begin{equation}
    c=(\rho,\psi,e,\mu)\in\mathcal{C},
    \label{eq:sga_configuration}
\end{equation}
where $\rho$ controls gradient retention, $\psi$ smooths local variations before structural extraction, $e$ selects central-difference edges, and $\mu$ measures correspondence between the resulting structural maps. $\rho$ applies only to the gradient branch, while $\psi$ and $e$ apply identically to both. All protocols share the scoring pipeline in \cref{fig:method} and differ only in configuration selection (\cref{sec:configuration_selection}).

\subsection{Structural Sensitivity Formulation}
\label{sec:structural_sensitivity_formulation}

SGA builds on a \emph{structural sensitivity hypothesis}: robust models tend to align local input sensitivity with stable input structure, whereas less robust models exhibit more incidental or fragmented sensitivity. This hypothesis defines a measurable property, not a certified robustness guarantee.

\textbf{Objective-conditioned local sensitivity}
For input $x$, conditioning output $\tau$, and differentiable objective $\mathcal{L}_f(x;\tau)$, define the local sensitivity field
\begin{equation}
    g_f(x;\tau)
    =
    \nabla_x \mathcal{L}_f(x;\tau).
    \label{eq:formal_sensitivity}
\end{equation}
For a sufficiently small perturbation $\delta$,
\begin{equation}
    \mathcal{L}_f(x+\delta;\tau)
    =
    \mathcal{L}_f(x;\tau)
    +
    \left\langle g_f(x;\tau),\delta\right\rangle
    +
    o(\|\delta\|).
    \label{eq:formal_local_response}
\end{equation}
Thus, $g_f(x;\tau)$ describes the first-order direction and spatial distribution of input influence on the conditioning objective. SGA characterizes this field through its spatial organization rather than its raw magnitude.

\textbf{Common spatial-structure domain}
The input image and its sensitivity field belong to different signal spaces and therefore are not directly comparable. We instead define the common binary spatial-structure domain
\begin{equation}
    \mathcal{S}_{h_x,w_x}
    =
    \left\{
        B \mid B\in\{0,1\}^{h_x\times w_x}
    \right\},
    \label{eq:formal_structural_domain}
\end{equation}
whose elements represent the spatial support of salient local variation on the image plane. $h_x$ and $w_x$ denote the spatial height and width of $x$. For configuration $c$, let
\begin{equation}
    \Phi_x^c(x)\in\mathcal{S}_{h_x,w_x},
    \qquad
    \Phi_g^c(g_f(x;\tau))\in\mathcal{S}_{h_x,w_x}
    \label{eq:formal_structural_mappings}
\end{equation}
denote the input-structure and sensitivity-structure mappings, respectively. They need not preserve original signal values but express both signals through the same spatial relation: where structurally salient variation occurs.

\textbf{Structural sensitivity alignment}
Given the similarity measure $\mu$ specified by configuration $c$, define the per-probe structural sensitivity alignment as
\begin{equation}
    A_c(f,x;\tau)
    =
    \mu\!\left(
        \Phi_x^c(x),
        \Phi_g^c(g_f(x;\tau))
    \right).
    \label{eq:formal_alignment}
\end{equation}
A larger value indicates stronger spatial correspondence between input structure and objective-conditioned sensitivity under $c$. SGA defines $\Phi_x^c$, $\Phi_g^c$, and $\mu$ below and aggregates $A_c$ over probes into a model-level score. Its link to adversarial robustness is empirically assessed against attack-based reference orders, not assumed by definition. We test spatial organization by permuting gradient locations while preserving non-spatial statistics (\cref{sec:ablation-spatial-destruction}).

\subsection{Objective Construction and Input-gradient Extraction}
\label{sec:gradient_extraction}

For each probe, SGA constructs an objective solely for input-gradient extraction. As the evaluated models expose different output interfaces, we define the conditioning output and corresponding loss separately for classification and VQA. For DNNs and VLMs evaluated on ImageNet-1K classification, we define the conditioning class as
\begin{equation}
    \widetilde{y}_{i,n}
    =
    \begin{cases}
        y_n,
        & \text{if $x_n$ is a test-set probe}, \\[2pt]
        \displaystyle\arg\max_y f_i(x_n)_y,
        & \text{if $x_n$ is a public probe},
    \end{cases}
    \label{eq:probe_target}
\end{equation}
where $y_n$ is the ground-truth class label.

For autoregressive LVLMs evaluated on VQA, we define the conditioning answer sequence as
\begin{equation}
    \widetilde{\mathbf{y}}_{i,n}
    =
    \begin{cases}
        \mathbf{y}_n,
        & \text{if $x_n$ is a test-set probe}, \\[2pt]
        \operatorname{Dec}^{\mathrm{greedy}}_{f_i}(x_n,q_n),
        & \text{if $x_n$ is a public probe},
    \end{cases}
    \label{eq:lvlm_probe_target}
\end{equation}
where $\mathbf{y}_n$ is the ground-truth answer sequence and $\operatorname{Dec}^{\mathrm{greedy}}_{f_i}(x_n,q_n)$ denotes the answer sequence greedily decoded by $f_i$ from image $x_n$ and VQA question prompt $q_n$.

Thus, test-set probes use the available ground-truth outputs, whereas public probes use outputs generated by each model. Public-probe evaluation consequently incurs an additional inference for output generation.

For DNNs and VLMs evaluated on ImageNet-1K classification, we use the cross-entropy loss
\begin{equation}
    \mathcal{L}_{i,n}
    =
    -\log p_{f_i}
    \left(
        \widetilde{y}_{i,n}\mid x_n
    \right).
    \label{eq:cls_loss}
\end{equation}

For autoregressive LVLMs evaluated on VQA, let
\begin{equation}
    \widetilde{\mathbf{y}}_{i,n}
    =
    \left(
        \widetilde{y}_{i,n,1},
        \ldots,
        \widetilde{y}_{i,n,T_{i,n}}
    \right)
\end{equation}
denote the conditioning answer sequence in \cref{eq:lvlm_probe_target}. We compute its teacher-forced mean negative log likelihood as
\begin{equation}
\begin{aligned}
    \mathcal{L}_{i,n}
    =
    -\frac{1}{T_{i,n}}
    \sum_{t=1}^{T_{i,n}}
    \log p_{f_i}\Bigl(
        \widetilde{y}_{i,n,t}
        \,\Big|\,
        \widetilde{\mathbf{y}}_{i,n,<t},
        x_n,
        q_n
    \Bigr).
\end{aligned}
    \label{eq:lvlm_loss}
\end{equation}

For either model interface, the input gradient is
\begin{equation}
    g_{i,n}
    =
    \nabla_{x_n}\mathcal{L}_{i,n}.
    \label{eq:input_gradient}
\end{equation}
Computing $g_{i,n}$ requires one backward pass per model--probe pair. Public-probe output generation adds inference but no additional backpropagation.

\subsection{Model Sensitivity Representation}
\label{sec:gradient_representation}

Raw gradient magnitudes are not directly comparable across model interfaces or architectures. Logit scale, answer length, preprocessing, and model parameterization can rescale $g_{i,n}$ without changing the spatial organization of its sensitivity. SGA therefore represents the gradient using retained signs, with the retained coordinates determined by
\begin{equation}
    \rho
    \in
    \left\{
        \operatorname{abs}(\tau_g),
        \operatorname{top}(p),
        \operatorname{all}
    \right\}.
    \label{eq:retention_rules}
\end{equation}
These rules respectively apply absolute thresholding, top-percentage retention, or no coordinate suppression. SGA adopts the resulting ternary sensitivity representation introduced by AdvSMs~\cite{wang2026shared}, using it as an intermediate representation for extracting within-model gradient structure.

Let $m_{\rho}(g)[u]\in\{0,1\}$ denote whether coordinate $u$ is retained under $\rho$. The ternary representation is
\begin{equation}
Q_{\rho}(g)[u]
=
\begin{cases}
    +1, & m_{\rho}(g)[u]=1 \ \text{and}\ g[u]>0,\\
    -1, & m_{\rho}(g)[u]=1 \ \text{and}\ g[u]<0,\\
    0, & \text{otherwise}.
\end{cases}
\label{eq:ternary}
\end{equation}
For $\rho=\operatorname{abs}(\tau_g)$,
\begin{equation}
    m_{\rho}(g)[u]
    =
    \mathbf{1}\!\left[|g[u]|\geq\tau_g\right].
\end{equation}
For $\rho=\operatorname{top}(p)$, the mask retains the $p$ percent largest values of $|g[u]|$ over all pixels and channels; for $\rho=\operatorname{all}$, every coordinate is retained.

We map $-1$, $0$, and $+1$ to grayscale levels $0$, $128$, and $255$, respectively, and merge RGB channels using luminance weights $(0.2989,0.5870,0.1140)$. The resulting representation preserves the spatial layout of positive, negative, and inactive sensitivity while reducing dependence on loss scale.

For $\rho=\operatorname{top}(p)$ and $\rho=\operatorname{all}$, the ternary representation is invariant to any positive rescaling of the gradient:
\begin{equation}
    Q_{\rho}(a g)=Q_{\rho}(g),
    \qquad a>0.
    \label{eq:positive_scale_invariance}
\end{equation}
Positive rescaling preserves gradient signs and, for $\operatorname{top}(p)$, the ordering of magnitudes. The absolute-threshold rule $\operatorname{abs}(\tau_g)$ does not have this invariance because its retained set depends on absolute scale.

\subsection{Structural Map Extraction}
\label{sec:structural_maps}

SGA applies the same structural extraction process to the input image and the model sensitivity representation. The input image is first converted to grayscale using the same luminance weights. SGA then applies the same smoothing transformation $\psi$ to the grayscale input and the channel-merged ternary sensitivity representation. The transformation $\psi$ can be the identity operation, Gaussian smoothing with standard deviation $\sigma$, or bilinear scale smoothing that downsamples by an integer factor $d$ and then restores the original resolution. The two non-identity smoothing operators are mutually exclusive. Smoothing suppresses fine-scale variation while retaining the regional organization of both representations.

SGA converts the smoothed representations into the common spatial-structure domain $\mathcal{S}_{h_x,w_x}$ defined in \cref{eq:formal_structural_domain}. Let $z$ denote either smoothed grayscale representation. At interior pixels, we compute horizontal and vertical responses via central differences, with boundary responses set to zero:
\begin{equation}
\begin{aligned}
    D_x z[u,v] &= z[u,v+1]-z[u,v-1],\\
    D_y z[u,v] &= z[u+1,v]-z[u-1,v].
\end{aligned}
\label{eq:central_differences}
\end{equation}
The edge magnitude is
\begin{equation}
    E(z)=\sqrt{(D_xz)^2+(D_yz)^2}.
\end{equation}
Let $T_e(z)$ be the $e$-th percentile among the nonzero values of $E(z)$. The corresponding binary structural map is
\begin{equation}
    B_e(z)[u,v]
    =
    \mathbf{1}\!\left[E(z)[u,v]\geq T_e(z)\right].
    \label{eq:edge}
\end{equation}
If $E(z)$ contains no nonzero values, $B_e(z)$ is defined as the all-zero map. Writing $\operatorname{Gray}(x_n)$ for grayscale conversion of the probe image and $Z_{i,n,\rho}$ for the retained, ternary-encoded, and channel-merged representation of $g_{i,n}$ under retention rule $\rho$, we obtain
\begin{equation}
    B^x_{n,c}
    =
    B_e\!\left(\psi\!\left(\operatorname{Gray}(x_n)\right)\right),
    \qquad
    B^g_{i,n,c}
    =
    B_e\!\left(\psi\!\left(Z_{i,n,\rho}\right)\right).
    \label{eq:structural_maps}
\end{equation}
These maps instantiate $\Phi_x^c(x_n)$ and $\Phi_g^c(g_{i,n})$ in \cref{eq:formal_structural_mappings} under the same smoothing and edge criteria.

\subsection{SGA Score}
\label{sec:sga_score}

SGA instantiates the structural sensitivity alignment in \cref{eq:formal_alignment} by comparing the input and sensitivity structural maps using Dice similarity, structural similarity (SSIM), or their arithmetic mean. Dice similarity is
\begin{equation}
    \operatorname{Dice}(A,B)
    =
    \frac{2|A\cap B|}
    {|A|+|B|+\varepsilon_{\mathrm{D}}},
    \label{eq:dice}
\end{equation}
where $\varepsilon_{\mathrm{D}}=10^{-12}$ is a numerical stabilizer. For binary maps, Dice has the exact mismatch form
\begin{equation}
    1-\operatorname{Dice}(A,B)
    =
    \frac{|A\triangle B|+\varepsilon_{\mathrm{D}}}
    {|A|+|B|+\varepsilon_{\mathrm{D}}},
    \label{eq:dice_mismatch}
\end{equation}
where $A\triangle B$ denotes the symmetric difference between their structural supports. Thus, Dice mismatch measures unsupported sensitivity structure and input structure lacking corresponding sensitivity, normalized by their total support. A derivation is provided in Appendix~\ref{app:fixed_configuration_proofs}. SSIM instead measures local luminance-and-structure agreement using the standard local-window formulation~\cite{wang2004image}:

\begin{equation}
\begin{aligned}
    L_w(A,B)
    &=
    \frac{2m_{A,w}m_{B,w}+C_1}
    {m_{A,w}^{2}+m_{B,w}^{2}+C_1},
    \\[3pt]
    V_w(A,B)
    &=
    \frac{2\kappa_{AB,w}+C_2}
    {v_{A,w}+v_{B,w}+C_2},
    \\[3pt]
    \operatorname{SSIM}(A,B)
    &=
    \frac{1}{|\mathcal{W}|}
    \sum_{w\in\mathcal{W}}
    L_w(A,B)V_w(A,B).
\end{aligned}
\label{eq:ssim}
\end{equation}
Here, $\mathcal{W}$ denotes local windows; $m_{A,w}$ and $m_{B,w}$ are local means; $v_{A,w}$ and $v_{B,w}$ are local variances; $\kappa_{AB,w}$ is local covariance; and $C_1$ and $C_2$ are stabilizers.

For the similarity measure $\mu$ specified by $c$, the SGA score of model $f_i$ on $K$ probes is
\begin{equation}
    S_c(f_i;\mathcal{D}_K)
    =
    \frac{1}{K}
    \sum_{n=1}^{K}
    \mu\!\left(B^x_{n,c},B^g_{i,n,c}\right).
    \label{eq:sga}
\end{equation}
The resulting continuous model-level SGA score requires no cross-model normalization, with larger values indicating stronger structural sensitivity alignment. We evaluate its utility as a robustness signal through agreement with the reference order. Within a homogeneous pool, the scores may additionally induce a descending model order, whose agreement with the reference order can be evaluated using $\mathrm{PA}$ from \cref{eq:pa}.

\subsection{Statistical Properties}
\label{sec:fixed_configuration_properties}

We analyze SGA's statistical behavior under a configuration $c$ fixed independently of the evaluation probes. Let $\mathcal{D}_1$ denote a randomly drawn single-probe batch under the construction in \cref{sec:probes}, including the associated question prompt in the LVLM setting, and let $\mathcal{P}$ denote its distribution. Define the population SGA score of model $f_i$ as
\begin{equation}
    \Theta_c(f_i;\mathcal{P})
    =
    \mathbb{E}_{\mathcal{D}_1\sim\mathcal{P}}
    \left[
        S_c(f_i;\mathcal{D}_1)
    \right].
    \label{eq:population_sga}
\end{equation}
If $\mathcal{D}_K$ is formed from $K$ independent single-probe draws according to $\mathcal{P}$, then $S_c(f_i;\mathcal{D}_K)$ in \cref{eq:sga} is the empirical estimator of \cref{eq:population_sga}.

\textbf{Pairwise concentration}
Suppose that, for fixed $c$, there exist finite $a_c<b_c$ such that
\begin{equation}
    S_c(f_i;\mathcal{D}_1)\in[a_c,b_c],
    \ \mathcal{P}\text{-a.s. for all }f_i\in\mathcal{M},
    \ r_c=b_c-a_c.
    \label{eq:bounded_sga_score}
\end{equation}
For any probe distribution $\mathcal{R}$, define the population score gap between two models as
\begin{equation}
    \Delta_{ij,c}^{\mathcal{R}}
    =
    \Theta_c(f_i;\mathcal{R})
    -
    \Theta_c(f_j;\mathcal{R}).
    \label{eq:population_sga_gap}
\end{equation}
If $\Delta_{ij,c}^{\mathcal{P}}>0$, the probability that $K$ probes reverse or tie the population SGA order of $f_i$ and $f_j$ is bounded by
\begin{equation}
    \Pr_{\mathcal{D}_K\sim\mathcal{P}^K}\![
        S_c(f_i;\mathcal{D}_K)
        \leq
        S_c(f_j;\mathcal{D}_K)
    ]
    \leq
    \exp\!\left(
        -\frac{
            K(\Delta_{ij,c}^{\mathcal{P}})^2
        }{
            2r_c^2
        }
    \right).
    \label{eq:pairwise_sga_concentration}
\end{equation}
Thus, recovery of the population SGA order for a model pair becomes more reliable with additional probes and larger population score gaps.

\textbf{Probe-distribution order preservation}
For two probe distributions $\mathcal{P}$ and $\mathcal{Q}$, define
\begin{equation}
    \operatorname{Disc}_c
    (\mathcal{P},\mathcal{Q};\mathcal{M})
    =
    \sup_{f_i\in\mathcal{M}}
    \left|
        \Theta_c(f_i;\mathcal{P})
        -
        \Theta_c(f_i;\mathcal{Q})
    \right|.
    \label{eq:sga_distribution_discrepancy}
\end{equation}
For every model pair,
\begin{equation}
    \left|
        \Delta_{ij,c}^{\mathcal{P}}
        -
        \Delta_{ij,c}^{\mathcal{Q}}
    \right|
    \leq
    2\operatorname{Disc}_c
    (\mathcal{P},\mathcal{Q};\mathcal{M}).
    \label{eq:sga_gap_transfer}
\end{equation}
Consequently, the SGA order of $f_i$ and $f_j$ under $\mathcal{P}$ is preserved under $\mathcal{Q}$ whenever
\begin{equation}
    \Delta_{ij,c}^{\mathcal{P}}
    >
    2\operatorname{Disc}_c
    (\mathcal{P},\mathcal{Q};\mathcal{M}).
    \label{eq:sga_order_preservation}
\end{equation}

For a configuration fixed independently of probes, empirical SGA pairwise order concentrates around its population counterpart, while population order is preserved across probe distributions if the score gap exceeds twice the distributional discrepancy. Proofs appear in Appendix~\ref{app:fixed_configuration_proofs}.

\subsection{Configuration Selection Protocols}
\label{sec:configuration_selection}

SGA uses the same hierarchical configuration-selection mechanism across all protocols. At each layer, only the configurations that maximize the corresponding criterion are retained. The layers follow the fixed order
\begin{equation}
    \mathrm{PA}
    \;\rightarrow\;
    \operatorname{Std}
    \;\rightarrow\;
    \operatorname{Gap}
    \;\rightarrow\;
    \text{seeded random}.
    \label{eq:selection_cascade}
\end{equation}
$\mathrm{PA}$ is the primary criterion and is applied first whenever a reference order is available, but its discrete nature can produce many ties. Score standard deviation is therefore the next criterion; in our experiments, it resolves all remaining ties after $\mathrm{PA}$ and yields a unique selection when $\mathrm{PA}$ is unavailable. The minimum adjacent gap and seeded random choice are retained only as fallback tie-breakers.

For any model set $\mathcal{A}$ and probe set $\mathcal{D}$, define the score vector produced by configuration $c\in\mathcal{C}$ as
\begin{equation}
    \mathbf{S}_c(\mathcal{A};\mathcal{D})
    =
    \left(
        S_c(f;\mathcal{D})
    \right)_{f\in\mathcal{A}}.
    \label{eq:configuration_score_vector}
\end{equation}
The score standard deviation is
\begin{equation}
    \operatorname{Std}_c(\mathcal{A};\mathcal{D})
    =
    \operatorname{Std}\!\left(
        \mathbf{S}_c(\mathcal{A};\mathcal{D})
    \right).
    \label{eq:configuration_std}
\end{equation}
To define the minimum adjacent score gap, let
\begin{equation}
    s_{c,(1)}
    \leq
    s_{c,(2)}
    \leq
    \cdots
    \leq
    s_{c,(|\mathcal{A}|)}
\end{equation}
denote the sorted entries of
$\mathbf{S}_c(\mathcal{A};\mathcal{D})$. Then,
\begin{equation}
    \operatorname{Gap}_c(\mathcal{A};\mathcal{D})
    =
    \min_{1\leq j<|\mathcal{A}|}
    \left(
        s_{c,(j+1)}-s_{c,(j)}
    \right).
    \label{eq:configuration_gap}
\end{equation}
Without a reference order, configuration selection aims to distinguish models more clearly through their structural-alignment scores. Score standard deviation captures overall separation, while the minimum adjacent score gap serves as a fallback tie-breaker for neighboring models. This reference-free criterion emphasizes the discriminative capacity of structural alignment itself, rather than agreement with an externally defined ranking.

When a reference $\mathbf{R}_{\mathcal{A}}$ is available, PA is computed as
\begin{equation}
    \mathrm{PA}_c(
        \mathcal{A},
        \mathcal{D};
        \mathbf{R}_{\mathcal{A}}
    )
    =
    \mathrm{PA}\!\left(
        \mathbf{S}_c(\mathcal{A};\mathcal{D}),
        \mathbf{R}_{\mathcal{A}}
    \right).
    \label{eq:configuration_pa}
\end{equation}
Only the order represented by $\mathbf{R}_{\mathcal{A}}$ is used; its numerical magnitudes are irrelevant.

For an ordered criterion sequence $\Gamma$, define the common configuration-selection rule as
\begin{equation}
    c_{\Gamma}^*
    (\mathcal{A},\mathcal{D};\mathbf{R}_{\mathcal{A}})
    =
    \operatorname{SeededChoice}\bigl(
        \operatorname*{arg\,max}_{c\in\mathcal{C}}^{\mathrm{lex}}
        \Gamma_c
        (\mathcal{A},\mathcal{D};\mathbf{R}_{\mathcal{A}})
    \bigr),
    \label{eq:configuration_selection}
\end{equation}
where $\Gamma$ is an order-preserving subsequence of
\begin{equation}
    \left(
        \mathrm{PA},
        \operatorname{Std},
        \operatorname{Gap}
    \right),
\end{equation}
and $\Gamma_c$ contains the corresponding criterion values for configuration $c$. Lexicographic maximization applies the included criteria in their stated order, and $\operatorname{SeededChoice}$ resolves any tie remaining after all included criteria have been applied. When PA is not included, no reference order is required, and we write $\mathbf{R}_{\mathcal{A}}=\varnothing$.

\subsubsection{Transfer Protocol}

The transfer protocol selects a configuration once using a development model set that does not overlap the target models to which the configuration will be applied. Let $\mathcal{M}_{\mathrm{dev}}$ denote the development models, $\mathcal{D}_{\mathrm{dev}}$ the development probes, and $\mathbf{R}_{\mathrm{dev}}$ their reference robustness order. It uses the complete PA--standard-deviation--gap cascade:
\begin{equation}
    c_{\mathrm{tra}}^*
    =
    c_{(\mathrm{PA},\,\operatorname{Std},\,\operatorname{Gap})}^*
    \left(
        \mathcal{M}_{\mathrm{dev}},
        \mathcal{D}_{\mathrm{dev}};
        \mathbf{R}_{\mathrm{dev}}
    \right).
    \label{eq:transfer_selection}
\end{equation}
The resulting configuration is frozen before target evaluation and applied to every target model $f\in\mathcal{M}$:
\begin{equation}
    S_{c_{\mathrm{tra}}^*}(f;\mathcal{D}_K).
    \label{eq:transfer_target_score}
\end{equation}
Neither the target-pool scores nor its reference robustness order participates in configuration selection. Once the transferred configuration has been fixed, each target model can therefore be scored independently.

\subsubsection{Uncalibrated Protocol}

The uncalibrated protocol assumes no reference robustness values or ordering. For each probe batch $\mathcal{D}_K$, it evaluates the candidate configurations on the complete target pool $\mathcal{M}$. Because PA cannot be computed without a reference order, this protocol uses only the standard-deviation and gap layers:
\begin{equation}
    c_{\mathrm{unc}}^*
    =
    c_{(\operatorname{Std},\,\operatorname{Gap})}^*
    \left(
        \mathcal{M},
        \mathcal{D}_K;
        \varnothing
    \right).
    \label{eq:uncalibrated_selection}
\end{equation}
The selected configuration is shared by all models in $\mathcal{M}$ for the current probe batch:
\begin{equation}
    \left\{
        S_{c_{\mathrm{unc}}^*}(f;\mathcal{D}_K)
        :
        f\in\mathcal{M}
    \right\}.
    \label{eq:uncalibrated_target_scores}
\end{equation}
Although this protocol requires no robustness reference, both its selected configuration and resulting scores depend on the composition of the target pool.

\subsubsection{Calibrated Protocol}

Let
\begin{equation}
    \mathcal{M}_{\mathrm{anc}}
    =
    \{f_1,\ldots,f_H\}
\end{equation}
denote the $H$ anchors assigned to the current probe batch, and let $\mathbf{R}_{\mathrm{anc}}$ denote their known robustness order. The remaining models form the target set $\mathcal{M}_{-\mathrm{anc}}$. For each probe batch $\mathcal{D}_K$, the calibrated protocol applies the complete cascade to the anchor models:
\begin{equation}
    c_{\mathrm{cal}}^*
    =
    c_{(\mathrm{PA},\,\operatorname{Std},\,\operatorname{Gap})}^*
    \left(
        \mathcal{M}_{\mathrm{anc}},
        \mathcal{D}_K;
        \mathbf{R}_{\mathrm{anc}}
    \right).
    \label{eq:calibrated_selection}
\end{equation}
The configuration is then shared by all targets:
\begin{equation}
    \left\{
        S_{c_{\mathrm{cal}}^*}(f;\mathcal{D}_K)
        :
        f\in\mathcal{M}_{-\mathrm{anc}}
    \right\}.
    \label{eq:calibrated_target_scores}
\end{equation}

Configuration selection uses only anchor scores and order, excluding target scores and reference order. To evaluate ranking agreement, we include all pairs with at least one target and exclude anchor--anchor pairs used for selection. The effect of anchor count $H$ is evaluated in \cref{sec:ablation_anchor}.

\subsubsection{Oracle Protocol}

The oracle protocol evaluates candidate configurations using the complete target pool $\mathcal{M}$ and its reference robustness order $\mathbf{R}$. For each probe batch $\mathcal{D}_K$, it applies the complete cascade:
\begin{equation}
    c_{\mathrm{ora}}^*
    =
    c_{(\mathrm{PA},\,\operatorname{Std},\,\operatorname{Gap})}^*
    \left(
        \mathcal{M},
        \mathcal{D}_K;
        \mathbf{R}
    \right).
    \label{eq:oracle_selection}
\end{equation}
The selected configuration is shared by the target pool:
\begin{equation}
    \left\{
        S_{c_{\mathrm{ora}}^*}(f;\mathcal{D}_K)
        :
        f\in\mathcal{M}
    \right\}.
    \label{eq:oracle_target_scores}
\end{equation}
Oracle SGA uses the complete target pool and its reference order, so is an analysis protocol rather than a deployable evaluation setting.

\subsection{Computational Complexity}
\label{sec:method_complexity}

SGA runtime includes model loading, conditioning-output inference, gradient computation and structural post-processing:
\begin{equation}
    T_{\mathrm{SGA}}
    =
    T_{\mathrm{load}}
    +
    T_{\mathrm{inf}}
    +
    T_{\mathrm{grad}}
    +
    T_{\mathrm{post}}.
    \label{eq:sga_cost_decomposition}
\end{equation}

Let $M=|\mathcal{M}|$ be the number of evaluated models, $K$ the number of probes, $C=|\mathcal{C}|$ the number of candidate configurations, and $P$ the number of spatial entries processed after channel merging. Let $L_i$, $I_i$, and $G_i$ denote the costs of loading model $f_i$, performing one inference, and computing one input gradient, respectively. Here, $G_i$ includes the forward evaluation of the gradient objective and its backward pass.

Models are loaded once and reused for every probe. Thus,
\begin{equation}
    T_{\mathrm{load}}
    =
    O\!\left(
        \sum_{i=1}^{M} L_i
    \right).
    \label{eq:loading_complexity}
\end{equation}
When a probe has no ground-truth output, one additional inference is required to generate its conditioning output. The corresponding inference cost is
\begin{equation}
    T_{\mathrm{inf}}
    =
    O\!\left(
        K\sum_{i=1}^{M} I_i
    \right).
    \label{eq:inference_complexity}
\end{equation}
Because each model--probe pair requires only one input-gradient computation,
\begin{equation}
    T_{\mathrm{grad}}
    =
    O\!\left(
        K\sum_{i=1}^{M} G_i
    \right).
    \label{eq:gradient_complexity}
\end{equation}

All candidate configurations reuse the same cached input gradients and require no additional model loading, inference, or backpropagation. With fixed-size smoothing and similarity windows, structural scoring is linear in $P$, giving at most $O(CMKP)$ operations across all models, probes, and configurations. In the worst case, configuration selection additionally requires pairwise PA computation and score sorting, bounded by $O(CM^2)$ and $O(CM\log M)$, respectively. Therefore,
\begin{equation}
    T_{\mathrm{post}}
    \leq
    T_{\mathrm{post}}^{\mathrm{worst}}
    =
    O\!\left(
        CMKP
        +
        CM^2
        +
        CM\log M
    \right).
    \label{eq:post_complexity}
\end{equation}

In our setting, $C$, $M$, and $K$ are small, while model loading, inference, and input-gradient computation require loading and executing deep models. Thus,
\begin{equation}
    T_{\mathrm{post}}
    \ll
    T_{\mathrm{load}}
    +
    T_{\mathrm{inf}}
    +
    T_{\mathrm{grad}},
    \label{eq:post_negligible}
\end{equation}
so the total runtime is dominated by the first three stages.

The transfer protocol applies only one fixed configuration and performs no configuration selection during target evaluation. Its post-processing cost is
\begin{equation}
    T_{\mathrm{post}}^{\mathrm{transfer}}
    =
    O(MKP).
    \label{eq:transfer_post_complexity}
\end{equation}
Compared with transfer SGA, the uncalibrated, calibrated, and oracle protocols add at most $0.032$ seconds. Their configuration-selection cost is therefore negligible relative to model loading, inference, and input-gradient computation.

\section{Evaluation}

\subsection{Experimental Settings}
\label{sec:experiments}

\textbf{Data and Model Pools}
For classification, we use the 1,000-image NIPS development set derived from ImageNet-1K~\cite{nips2017} as test-set probes. For VQA, we use VQAv2 question--image pairs with COCO images~\cite{antol2015vqa,lin2014microsoft} as test-set probes. For both tasks, we use CelebA-HQ~\cite{Lee2020mask} and Places365~\cite{zhou2017places} as public-image sources; for VQA, these public images are paired with VQAv2 questions without preserving the original image--question associations. We evaluate four threat-homogeneous pools containing 39 entries from 38 distinct model systems: 13 DNN $\ell_2$, 14 DNN $\ell_\infty$, seven VLM $\ell_\infty$, and five LVLM $\ell_\infty$ entries. All classification probes are processed at $224\times224$; the protocol-specific LVLM preprocessing is detailed in Appendix \ref{app:data-details}. The complete data construction and model composition appear in Appendices \ref{app:data-details} and \ref{app:models}.

\textbf{Reference Protocols}
We use PGD-50 as the primary reference for four pools: DNN $\ell_2$ uses $\epsilon=2$ and step size $0.1$, while three $\ell_\infty$ pools use $\epsilon=8/255$ and step size $1/255$, all with one restart and seed 0. We also use standard AutoAttack~\cite{croce2020reliable,croce2021robustbench} results from RobustBench at $\epsilon=4/255$ for DNN $\ell_\infty$. Reference scores and orders are computed separately within homogeneous pools. Distinct attack algorithms and budgets enable comparison across evaluation settings. Their protocol-dependent rankings serve as empirical references for SGA agreement, not ground truth. Appendix~\ref{app:models} details all protocols and values.

\textbf{Compared Methods and SGA Protocols}
We compare SGA with FGSM~\cite{goodfellow2014explaining}, GREAT~\cite{li2024great}, CLEVER~\cite{weng2018evaluating}, and Linearized Robustness~\cite{etmann2019connection}. Their exact configurations and LVLM adaptations are provided in Appendix \ref{app:baseline-details}. We instantiate the Transfer, Uncalibrated, Calibrated, and Oracle SGA protocols defined in \cref{tab:sga_protocols,sec:configuration_selection}. All four use the same 4,608-configuration candidate space and the selection cascade in \cref{eq:selection_cascade}; they differ only in the reference information permitted by their definitions. Transfer uses a single preset selected across all $K=10$ batches of a nonoverlapping development model set, whereas the other three protocols select one configuration for each probe batch. Transfer presets, candidate values, and the experimental realization of each protocol are given in Appendix \ref{app:configuration-selection}.

\begin{table*}[!t]
\centering
\tablefontsize
\begin{threeparttable}
\caption{\textbf{Pairwise agreement on unlabeled CelebA-HQ public probes at $K=10$.} Mean PA $\pm$ standard deviation across probe batches; best result in each model pool in bold.}
\label{tab:main-results-celeba}
\setlength{\tabcolsep}{5.6mm}
\begin{tabular}{llcccc}
\toprule
Method & Knowledge
& DNN $\ell_2$
& DNN $\ell_\infty$
& VLM $\ell_\infty$
& LVLM $\ell_\infty$ \\
\midrule
GREAT
& None
& $0.4326{\pm}0.1086$
& $0.4881{\pm}0.1216$
& $0.7586{\pm}0.1233$
& $0.4600{\pm}0.2332$ \\

FGSM
& $\ell_2$ or $\ell_\infty$
& $0.5113{\pm}0.1464$
& $0.3830{\pm}0.1440$
& $0.5714{\pm}0.1990$
& $0.5360{\pm}0.1676$ \\

CLEVER
& $\ell_2$ or $\ell_\infty$
& $0.8691{\pm}0.0399$
& $0.7977{\pm}0.0772$
& $0.7407{\pm}0.0590$
& $0.4060{\pm}0.1907$ \\

Linearized
& None
& $0.8649{\pm}0.0455$
& $0.7170{\pm}0.0721$
& $0.7800{\pm}0.0799$
& $0.6250{\pm}0.1228$ \\
\midrule
SGA (Uncalibrated)
& None
& $\mathbf{0.9694{\pm}0.0170}$
& $\mathbf{0.8651{\pm}0.0234}$
& $\mathbf{0.8010{\pm}0.0464}$
& $\mathbf{0.7620{\pm}0.1198}$ \\
\bottomrule
\end{tabular}
\end{threeparttable}
\end{table*}

\begin{table*}[!t]
\centering
\tablefontsize
\begin{threeparttable}
\caption{\textbf{Pairwise agreement on unlabeled Places365 public probes at $K=10$.} Mean PA $\pm$ standard deviation across probe batches; best result in each model pool in bold.}
\label{tab:main-results-places}
\setlength{\tabcolsep}{5.6mm}
\begin{tabular}{llcccc}
\toprule
Method & Knowledge
& DNN $\ell_2$
& DNN $\ell_\infty$
& VLM $\ell_\infty$
& LVLM $\ell_\infty$ \\
\midrule
GREAT
& None
& $0.4217{\pm}0.0992$
& $0.4665{\pm}0.1109$
& $0.4505{\pm}0.1235$
& $0.4440{\pm}0.2255$ \\

FGSM
& $\ell_2$ or $\ell_\infty$
& $0.5447{\pm}0.1564$
& $0.4316{\pm}0.1807$
& $0.2662{\pm}0.1954$
& $0.5820{\pm}0.1564$ \\

CLEVER
& $\ell_2$ or $\ell_\infty$
& $0.8949{\pm}0.0420$
& $0.8492{\pm}0.0579$
& $0.7262{\pm}0.0688$
& $0.4530{\pm}0.1947$ \\

Linearized
& None
& $0.8945{\pm}0.0428$
& $0.7966{\pm}0.0580$
& $0.7110{\pm}0.1039$
& $0.5580{\pm}0.1159$ \\
\midrule
SGA (Uncalibrated)
& None
& $\mathbf{0.9581{\pm}0.0308}$
& $\mathbf{0.8745{\pm}0.0365}$
& $\mathbf{0.7686{\pm}0.0415}$
& $\mathbf{0.6990{\pm}0.1245}$ \\
\bottomrule
\end{tabular}
\end{threeparttable}
\end{table*}

\begin{figure*}[!t]
    \centering
    
    \includegraphics[width=\textwidth]{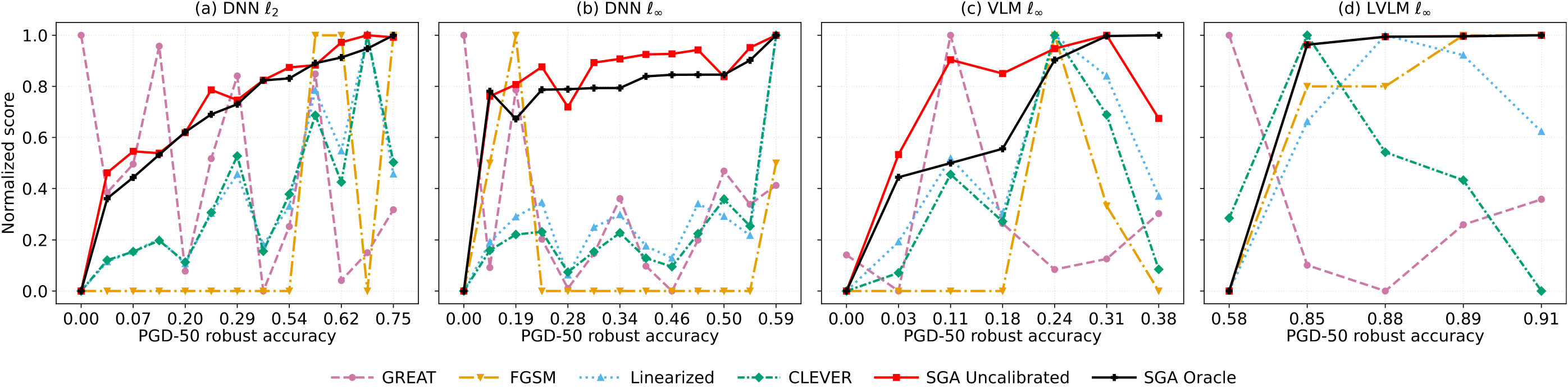}
    \caption{\textbf{Model scores versus the PGD-50 reference on CelebA-HQ public probes.} All methods use the same ten randomly selected probes, and scores are min--max normalized within each method and pool. Models are ordered by PGD-50 robust accuracy, so a more monotonically increasing curve indicates stronger ranking agreement. Uncalibrated SGA follows the reference order more consistently than the baselines and remains visually close to Oracle SGA.}
    \label{fig:agreement}
\end{figure*}

\textbf{Reporting Protocol}
For each $K\in\{1,2,5,10\}$, the fixed 1,000-example order is partitioned into consecutive, non-overlapping batches shared by all methods. We compute PA (defined in \cref{eq:pa}) separately for every batch and report its mean $\pm$ standard deviation across batches; the deviation measures sensitivity to probe-batch composition rather than model-training randomness. Unless stated otherwise, all main comparisons use $K=10$. Runtime is measured for the complete model pool and one $K=10$ batch on one NVIDIA H100 GPU with 94\,GB memory. Further batching, reporting, and timing details appear in Appendix \ref{app:runtime-details}.


\subsection{Results}
\label{sec:results}

\subsubsection{Main comparison}
\cref{tab:main-results-celeba,tab:main-results-places,fig:agreement} compare SGA with existing estimators in recovering attack-based robustness orders using ten unlabeled public probes under their respective knowledge settings.
\noindent \emph{i)} Uncalibrated SGA evaluates robustness effectively without evaluator knowledge. Across eight probe-source--model-pool conditions, it achieves an average PA of $0.8372$ with the same knowledge-free protocol for DNN and VLM classification and LVLM VQA.
\noindent \emph{ii)} Uncalibrated SGA outperforms the strongest overall baseline. It leads in all eight conditions, improving average PA by $9.38$ percentage points. \cref{fig:agreement} shows closer agreement with PGD-50 rankings than the baselines.
\noindent \emph{iii)} GREAT and FGSM perform poorly under few-shot public probing. Their average PAs are $0.4903$ and $0.4783$, respectively; even providing FGSM with the corresponding perturbation norm does not remedy this failure.
\noindent \emph{iv)} CLEVER and Linearized Robustness fail to transfer to generative VQA. Their average PAs are $0.8130$ and $0.7940$ across six classification conditions but drop to $0.4295$ and $0.5915$ on LVLMs, respectively, while SGA maintains $0.7305$ without protocol changes. Appendix~\ref{app:method-comparison} discusses the classifier-specific assumptions limiting their adaptation to open-ended LVLM outputs.

\begin{table*}[!t]
\centering
\tablefontsize
\caption{\textbf{Few-shot performance of Uncalibrated SGA on public probes.} Mean PA $\pm$ batch-level standard deviation.}
\label{tab:batch-size-results}
\setlength{\tabcolsep}{6.8mm}
\begin{tabular}{llcccc}
\toprule
Pool & Probe source & $K=10$ & $K=5$ & $K=2$ & $K=1$ \\
\midrule
\multirow{2}{*}{DNN $\ell_2$}
& CelebA-HQ
& $0.9694{\pm}0.0170$ & $0.9604{\pm}0.0216$
& $0.9400{\pm}0.0297$ & $0.9171{\pm}0.0381$ \\
& Places365
& $0.9581{\pm}0.0308$ & $0.9362{\pm}0.0468$
& $0.8989{\pm}0.0801$ & $0.8749{\pm}0.0952$ \\
\midrule
\multirow{2}{*}{DNN $\ell_\infty$}
& CelebA-HQ
& $0.8651{\pm}0.0234$ & $0.8538{\pm}0.0281$
& $0.8355{\pm}0.0386$ & $0.8089{\pm}0.0571$ \\
& Places365
& $0.8745{\pm}0.0365$ & $0.8484{\pm}0.0540$
& $0.7977{\pm}0.0936$ & $0.7409{\pm}0.1212$ \\
\midrule
\multirow{2}{*}{VLM $\ell_\infty$}
& CelebA-HQ
& $0.8010{\pm}0.0464$ & $0.7988{\pm}0.0470$
& $0.7890{\pm}0.0608$ & $0.7830{\pm}0.0747$ \\
& Places365
& $0.7686{\pm}0.0415$ & $0.7540{\pm}0.0567$
& $0.7371{\pm}0.0735$ & $0.7223{\pm}0.0867$ \\
\midrule
\multirow{2}{*}{LVLM $\ell_\infty$}
& CelebA-HQ
& $0.7620{\pm}0.1198$ & $0.7350{\pm}0.1307$
& $0.7280{\pm}0.1301$ & $0.7143{\pm}0.1365$ \\
& Places365
& $0.6990{\pm}0.1245$ & $0.7060{\pm}0.1219$
& $0.7078{\pm}0.1254$ & $0.7015{\pm}0.1286$ \\
\bottomrule
\end{tabular}
\end{table*}

\subsubsection{Few-shot performance}
\cref{tab:batch-size-results} evaluates Uncalibrated SGA with fewer public probes and no probe selection.
\noindent \emph{i)} Unlabeled public probes support few-shot robustness evaluation.
Average PA drops from $0.8372$ at $K=10$ to $0.7829$ at $K=1$, remaining informative with a single task-independent probe.
\noindent \emph{ii)} Random probe selection remains effective on average.
At $K=10$, the average batch-level standard deviation across eight conditions is $0.0550$, demonstrating effective average agreement without specialized probe selection.

\subsubsection{Comparison of SGA protocols}
\cref{tab:sga-protocols} compares three deployable SGA protocols and the Oracle upper bound using task-independent public probes.
\noindent \emph{i)} Transferred configurations remain effective despite limited generalization.
Transfer SGA achieves $0.8348$ average PA, $0.0024$ below Uncalibrated SGA, but exceeds it in only three of eight conditions, indicating modest overfitting and mismatch across evaluation settings.
\noindent \emph{ii)} Ordered anchors guide configuration selection.
Calibrated SGA outperforms Uncalibrated SGA in five of eight conditions, raising average PA from $0.8372$ to $0.8536$ with limited robustness-order information.
\noindent \emph{iii)} Configuration selection has substantial room for improvement.
Oracle SGA achieves $0.9782$ average PA, above all deployable protocols, demonstrating SGA's ability to closely match attack-based reference orders with suitable configurations while highlighting practical selection as an open challenge.

\begin{table}[!t]
\centering
\tablefontsize
\caption{\textbf{Comparison of SGA protocols at $K=10$.} Mean PA across probe batches.}
\label{tab:sga-protocols}
\setlength{\tabcolsep}{1.6mm}
\begin{tabular}{llcccc}
\toprule
Probe source & Protocol
& DNN $\ell_2$
& DNN $\ell_\infty$
& VLM $\ell_\infty$
& LVLM $\ell_\infty$ \\
\midrule

\multirow{4}{*}{CelebA-HQ}
& Uncalibrated & $0.9694$ & $0.8651$ & $0.8010$ & $0.7620$ \\
& Transfer     & $0.9592$ & $0.7910$ & $0.7700$ & $0.8320$ \\
& Calibrated   & $0.9667$ & $0.8515$ & $0.8233$ & $0.7600$ \\
& Oracle       & $0.9991$ & $0.9600$ & $0.9600$ & $0.9910$ \\
\midrule

\multirow{4}{*}{Places365}
& Uncalibrated & $0.9581$ & $0.8745$ & $0.7686$ & $0.6990$ \\
& Transfer     & $0.9426$ & $0.8137$ & $0.7981$ & $0.7720$ \\
& Calibrated   & $0.9767$ & $0.8846$ & $0.7883$ & $0.7775$ \\
& Oracle       & $0.9996$ & $0.9666$ & $0.9510$ & $0.9980$ \\
\bottomrule
\end{tabular}
\end{table}

\subsubsection{Sensitivity to the robustness reference}
\label{sec:reference-sensitivity}
\cref{tab:reference-comparison} examines how SGA agreement changes when switching the DNN $\ell_\infty$ reference from PGD-50 to AutoAttack using task-independent public probes.
\noindent \emph{i)} Agreement remains broadly stable across references. PGD-50 and AutoAttack achieve a reference-to-reference PA of $0.9670$. Switching references changes PA by only $0.0211$ on average across three deployable protocols and two probe sources, indicating stability across closely related attack-based orders. Additional references with similarly concordant rankings would offer limited new information; we therefore focus on the disputed model pairs to better understand the differences between these empirical robustness measurements.
\noindent \emph{ii)} Uncalibrated SGA favors AutoAttack on disputed pairs. The references disagree on only three model pairs (\cref{tab:reference-disagreement}). Despite selecting configurations independently of either reference, Uncalibrated SGA shows higher agreement with AutoAttack across both probe sources, driven entirely by these pairs.
\noindent \emph{iii)} This preference does not establish a uniquely correct reference. The PGD-50-tuned Oracle instead strongly favors PGD-50 on these pairs (\cref{tab:reference-disagreement}). This reversal shows that agreement depends on configuration selection rather than establishing either reference as definitive, motivating tests against more substantially different robustness benchmarks.

\begin{table}[!t]
\centering
\tablefontsize
\caption{\textbf{SGA agreement with PGD-50 and AutoAttack for DNN $\ell_\infty$.} Mean PA at $K=10$ using public probes.}
\label{tab:reference-comparison}
\setlength{\tabcolsep}{3.3mm}
\begin{tabular}{lcccc}
\toprule
& \multicolumn{2}{c}{CelebA-HQ}
& \multicolumn{2}{c}{Places365} \\
\cmidrule(lr){2-3}\cmidrule(lr){4-5}
Protocol
& PGD-50 & AutoAttack
& PGD-50 & AutoAttack \\
\midrule
Uncalibrated & $0.8651$ & $0.8974$ & $0.8745$ & $0.8886$ \\
Transfer     & $0.7910$ & $0.7705$ & $0.8137$ & $0.7821$ \\
Calibrated   & $0.8515$ & $0.8677$ & $0.8846$ & $0.8962$ \\
Oracle       & $0.9600$ & $0.9605$ & $0.9666$ & $0.9658$ \\
\bottomrule
\end{tabular}
\end{table}

\begin{table}[!t]
\centering
\tablefontsize
\caption{\textbf{SGA preferences on model pairs disputed by PGD-50 and AutoAttack.} Percentage of pair--batch comparisons supporting each reference order at $K=10$.}
\label{tab:reference-disagreement}
\setlength{\tabcolsep}{4.9mm}
\begin{tabular}{llcc}
\toprule
Probe source & Protocol
& PGD-50 & AutoAttack \\
\midrule
\multirow{2}{*}{CelebA-HQ}
& Uncalibrated & $1.0\%$  & $99.0\%$ \\
& Oracle       & $84.7\%$ & $15.3\%$ \\
\midrule
\multirow{2}{*}{Places365}
& Uncalibrated & $28.7\%$ & $71.3\%$ \\
& Oracle       & $90.7\%$ & $9.3\%$ \\
\bottomrule
\end{tabular}
\end{table}

\begin{table*}[!t]
\centering
\tablefontsize
\caption{\textbf{Runtime at $K=10$.} End-to-end seconds for baselines and Transfer SGA (test-set/public); other protocols report extra runtime.}
\label{tab:runtime}
\setlength{\tabcolsep}{8.7mm}
\begin{tabular}{lcccc}
\toprule
Method
& DNN $\ell_2$
& DNN $\ell_\infty$
& VLM $\ell_\infty$
& LVLM $\ell_\infty$ \\
\midrule
GREAT
& $6.6/7.1$
& $16.2/16.3$
& $19.0/19.2$
& $79.4/97.3$ \\

FGSM
& $7.5/7.9$
& $18.6/19.4$
& $21.3/22.7$
& $94.4/137.0$ \\

CLEVER
& $178.6/210.5$
& $218.3/239.0$
& $233.8/313.1$
& $1{,}215.2/1{,}811.3$ \\

Linearized
& $8.5/9.0$
& $23.1/27.0$
& $23.9/24.0$
& $113.2/118.1$ \\
\midrule

SGA (Transfer)
& $6.8/7.4$
& $17.9/18.2$
& $19.8/20.4$
& $84.6/103.8$ \\

SGA (Others)
& ${+}0.023/{+}0.025$
& ${+}0.023/{+}0.021$
& ${+}0.024/{+}0.024$
& ${+}0.032/{+}0.031$ \\
\bottomrule
\end{tabular}
\end{table*}

\subsubsection{Runtime}
\cref{tab:runtime} compares end-to-end runtimes.
\noindent \emph{i)} Low-cost methods have comparable runtimes. GREAT uses only forward inference, while FGSM, Linearized Robustness, and SGA require one input gradient per model--probe pair, yielding similarly low runtimes. CLEVER is much slower due to repeated sampling and optimization of local gradient norms. For comparison, AutoAttack takes about $5.4$ hours on 1,000 examples for one \texttt{Amini2024MeanSparse\_Swin-L} model, whereas SGA runtimes cover entire model pools at $K=10$.
\noindent \emph{ii)} Public probes incur modest inference overhead. Public probes require pseudo-target inference (\cref{sec:probes,sec:method_complexity}), increasing runtime, especially for larger LVLMs.
\noindent \emph{iii)} SGA configuration selection adds negligible overhead. Relative to Transfer SGA, other protocols add at most $0.032$ seconds by reusing cached gradients without extra backpropagation, making selection negligible relative to gradient extraction.


\subsection{Ablation Studies}
\label{sec:ablations}

We conduct ablation studies to examine factors affecting SGA's performance and behavior, following Appendix~\ref{app:ablation-protocols}.

\subsubsection{Spatial-Structure Destruction}
\label{sec:ablation-spatial-destruction}

This ablation tests whether SGA's robustness-ranking signal depends on spatial sensitivity organization beyond non-spatial gradient statistics. In \cref{tab:ablation-spatial-destruction}, disrupting spatial organization reduces PA across all eight pool--probe conditions by $0.2165$ on average, with consistently low variation across permutation seeds. This supports the structural sensitivity hypothesis (\cref{sec:structural_sensitivity_formulation}): spatial organization of objective-conditioned sensitivity provides substantial robustness-order information beyond gradient statistics alone. Residual agreement after permutation shows that spatial organization is important but not the sole source of SGA's ranking signal.

\begin{table}[!t]
\centering
\tablefontsize
\caption{\textbf{Spatial-structure destruction.} PA at $K=10$ using original and spatially permuted sensitivity fields. Permuted results are mean $\pm$ standard deviation over five seeds.}
\label{tab:ablation-spatial-destruction}
\setlength{\tabcolsep}{2.6mm}
\begin{tabular}{llccc}
\toprule
Probe source & Model pool & Original & Spatially permuted & $\Delta\mathrm{PA}$ \\
\midrule
\multirow{4}{*}{CelebA-HQ} & DNN $\ell_2$       & 0.9694 & $0.8877 \pm 0.0011$ & 0.0817 \\
           & DNN $\ell_\infty$  & 0.8651 & $0.7955 \pm 0.0031$ & 0.0696 \\
           & VLM $\ell_\infty$  & 0.8010 & $0.5647 \pm 0.0146$ & 0.2363 \\
           & LVLM $\ell_\infty$ & 0.7620 & $0.3350 \pm 0.0149$ & 0.4270 \\
\midrule
\multirow{4}{*}{Places365} & DNN $\ell_2$       & 0.9581 & $0.6909 \pm 0.0068$ & 0.2672 \\
           & DNN $\ell_\infty$  & 0.8745 & $0.7678 \pm 0.0012$ & 0.1067 \\
           & VLM $\ell_\infty$  & 0.7686 & $0.3854 \pm 0.0074$ & 0.3832 \\
           & LVLM $\ell_\infty$ & 0.6990 & $0.5390 \pm 0.0072$ & 0.1600 \\
\midrule
\multicolumn{2}{l}{Average}     & 0.8372 & 0.6208 & 0.2165 \\
\bottomrule
\end{tabular}
\end{table}

\subsubsection{Number of Anchors}
\label{sec:ablation_anchor}

\begin{figure}[!t]
    \centering
    \includegraphics[width=\columnwidth]{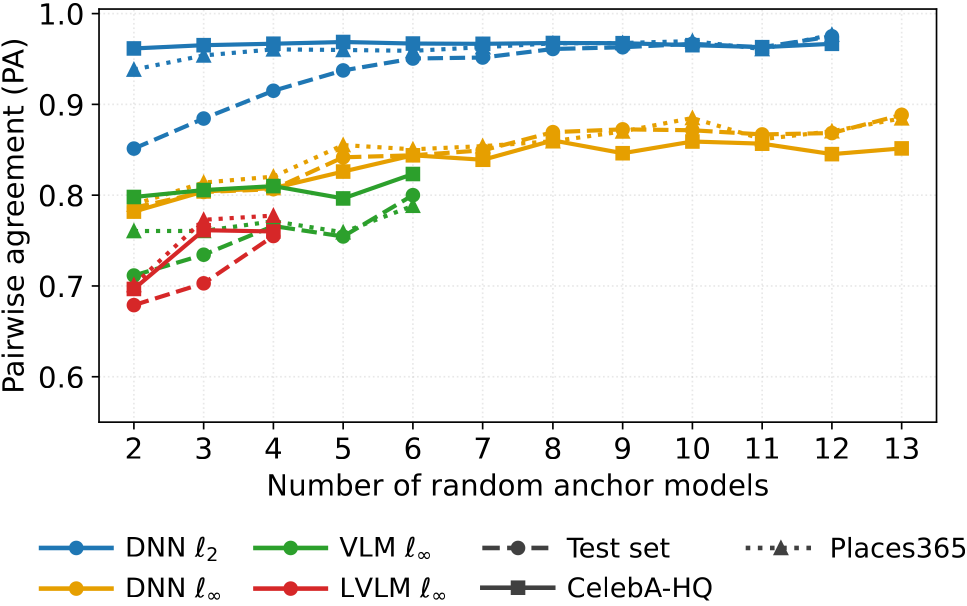}
    \caption{\textbf{Effect of the number of anchor models.} Using more randomly selected anchors generally improves PA, although the improvement becomes smaller as more anchors are added and varies across model pools.}
    \label{fig:ablation-anchor-num}
\end{figure}

This ablation examines anchor-set size in Calibrated SGA. \cref{fig:ablation-anchor-num} shows that PA generally rises with more anchors but with diminishing returns. Gains vary by pool: DNN $\ell_\infty$ benefits more consistently, while other pools saturate earlier or fluctuate modestly.

\subsubsection{Architecture Dependence}
\label{sec:ablation-architecture}

\begin{figure*}[!t]
    \centering
    \includegraphics[width=\textwidth]{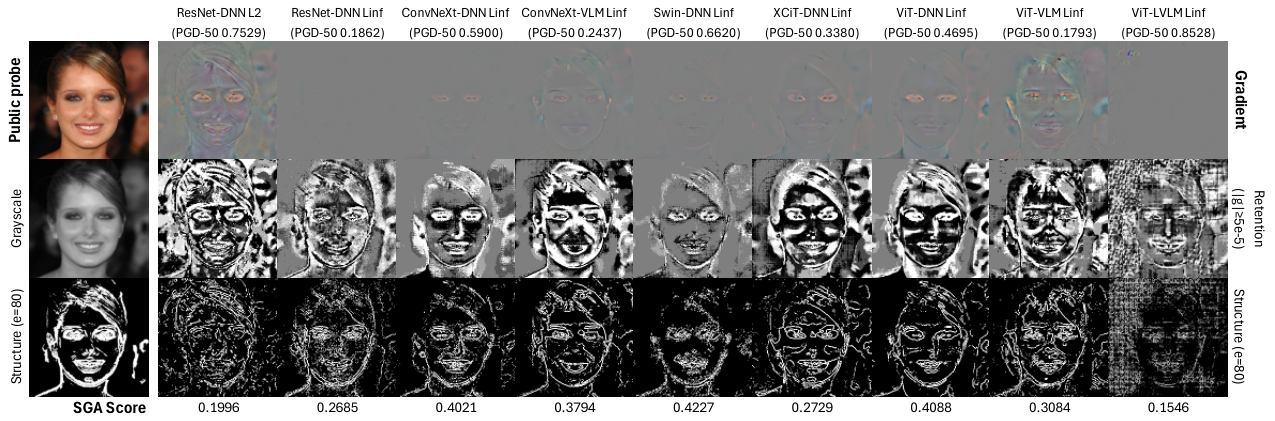}
    \caption{\textbf{Gradient-to-structure transformation across architectures.} Under the same input and configuration, ResNet, ConvNeXt, and Swin models exhibit relatively continuous sensitivity patterns, whereas the more challenging ViT and XCiT models exhibit more patchwise and fragmented patterns. Scores and PGD references are comparable only within pools.}
    \label{fig:visual-structure}
\end{figure*}

This ablation evaluates SGA's within-architecture ranking consistency, not cross-architecture comparisons. Across three probe sources, \cref{tab:architecture-ablation} shows higher PA for ResNet and ConvNeXt than ViT and XCiT with one fixed configuration per subgroup. Swin achieves similar PA, reflecting CNN-like locality from hierarchical shifted windows, while XCiT performs like ViT. Consistently, \cref{fig:visual-structure} shows continuous sensitivity patterns for ResNet, ConvNeXt, and Swin, versus patchwise, fragmented patterns for ViT and XCiT. These findings support the structural sensitivity hypothesis: SGA is more informative when objective-conditioned sensitivity exhibits coherent spatial organization in the common structural domain.

\begin{table}[!t]
\centering
\tablefontsize
\caption{\textbf{Architecture-level analysis across three data sources.} Best mean PA at $K=10$ within each architecture group, with configurations selected separately by group and data source. Numbers in parentheses indicate group sizes.}
\label{tab:architecture-ablation}
\setlength{\tabcolsep}{2.6mm}
\begin{tabular}{llccc}
\toprule
Architecture & Model pool & Test set & CelebA-HQ & Places365 \\
\midrule
ResNet (12)
& DNN $\ell_2$
& $0.9771$ & $0.9765$ & $0.9761$ \\
ResNet (2)
& DNN $\ell_\infty$
& $1.0000$ & $1.0000$ & $1.0000$ \\

ConvNeXt (3)
& DNN $\ell_\infty$
& $1.0000$ & $1.0000$ & $1.0000$ \\
ConvNeXt (2)
& VLM $\ell_\infty$
& $1.0000$ & $1.0000$ & $1.0000$ \\

Swin (3)
& DNN $\ell_\infty$
& $1.0000$ & $1.0000$ & $1.0000$ \\

ViT (2)
& DNN $\ell_\infty$
& $0.8900$
& $0.7800$
& $0.8000$ \\
ViT (5)
& VLM $\ell_\infty$
& $0.8783$
& $0.9083$
& $0.8950$ \\

XCiT (3)
& DNN $\ell_\infty$
& $0.9367$
& $0.8733$
& $0.9367$ \\

ViT + LLaVA (5)
& LVLM $\ell_\infty$
& $0.8850$
& $0.8850$
& $0.9370$ \\
\bottomrule
\end{tabular}
\end{table}

\subsubsection{Configuration Components}
\label{sec:ablation-configurations}

\begin{figure*}[!t]
    \centering
    \includegraphics[width=\textwidth]{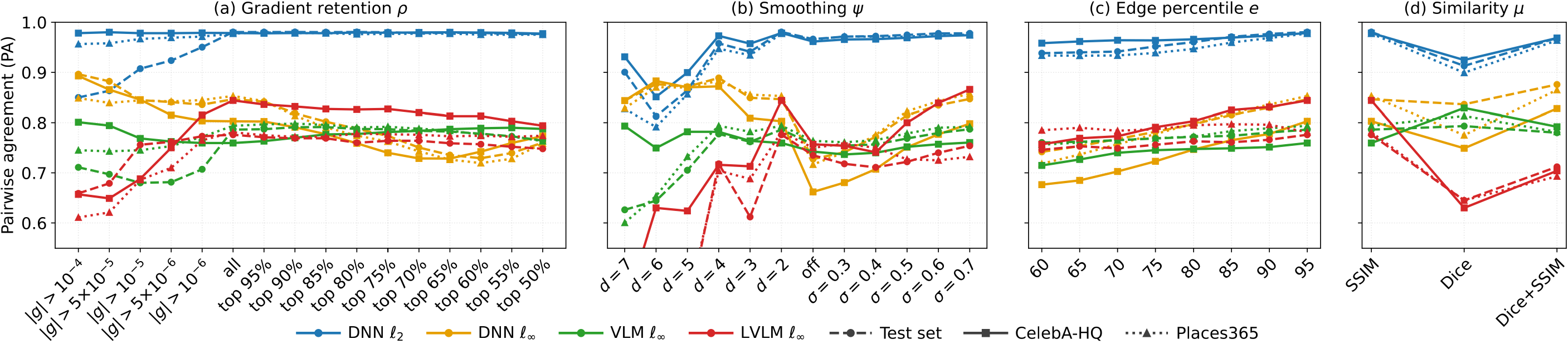}
    \caption{\textbf{Effects of the four SGA configuration components.} We vary one component at a time over its candidate range while fixing the other three to the global base configuration. Each point reports the mean PA across all probe batches at $K=10$.}
    \label{fig:ablation-configs}
\end{figure*}

This ablation varies each configuration component around a global base to assess consistency across pools and probe sources. In \cref{fig:ablation-configs}, edge percentile shows the clearest trend: increasing it from 60 to 95 raises mean PA from $0.7905$ to $0.8483$, with 95 best in 11 of 12 conditions. Smoothing also helps: $d=2$ outperforms no smoothing in all conditions and achieves the highest mean PA, although optimal strength varies and excessive downscaling can sharply reduce PA. Gradient retention and similarity metric are more pool-dependent. Retaining all coordinates or the top 95\% yields the highest mean PAs ($0.8483$ and $0.8461$), while stronger filtering helps some pools but harms others. Metric preferences are consistent across probe sources within each pool. Thus, some effects generalize, while others depend on the model pool.

\subsubsection{Individual Probes}
\label{sec:ablation-probes}
\begin{figure}[!t]
    \centering
    \includegraphics[width=\linewidth]{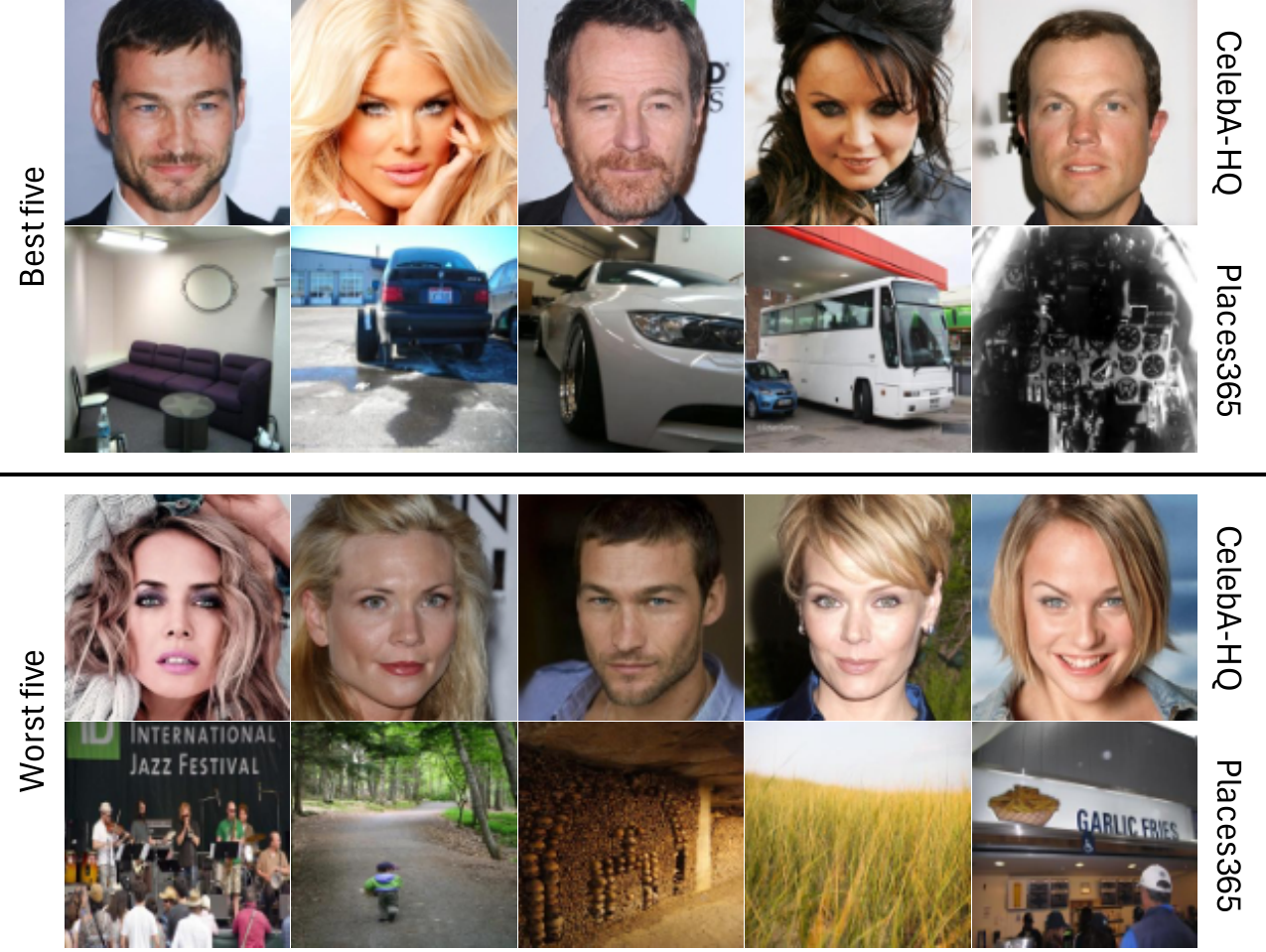}
    \caption{\textbf{Variation across individual probes.} The five highest- and five lowest-ranked probes from CelebA-HQ and Places365.}
    \label{fig:ablation-samples}
\end{figure}

This ablation examines whether visual properties distinguish probes that are easier or harder for SGA. In \cref{tab:ablation-samples}, the five best probes exceed the five worst by $0.1083$ on CelebA-HQ and $0.0518$ on Places365 on average. However, the ordering reverses for CelebA-HQ VLMs and ties for Places365 VLMs and LVLMs, showing that probe suitability varies across pools. \cref{fig:ablation-samples} shows no reliable visual distinction between the best and worst probes. Both CelebA-HQ groups contain similarly centered portraits, while the worst Places365 probes include homogeneous and cluttered images. CelebA-HQ achieves $0.0635$ higher average PA than Places365 across these groups, possibly due to more standardized facial composition versus the varied layouts and structural complexity of Places365. However, no reliable visual probe-selection rule emerges.

\begin{table}[!t]
\centering
\tablefontsize
\caption{\textbf{Variation across individual probes.} PA of the $K=5$ batches formed from the five highest- and five lowest-ranked probes in each public-probe source.}
\label{tab:ablation-samples}
\setlength{\tabcolsep}{5.2mm}
\begin{tabular}{llcc}
\toprule
Probe source & Model pool & Best five & Worst five \\
\midrule
\multirow{5}{*}{CelebA-HQ}
& DNN $\ell_2$       & $0.9744$ & $0.9487$ \\
& DNN $\ell_\infty$  & $0.9011$ & $0.8462$ \\
& VLM $\ell_\infty$  & $0.7619$ & $0.8095$ \\
& LVLM $\ell_\infty$ & $1.0000$ & $0.6000$ \\
& Average            & $0.9094$ & $0.8011$ \\
\midrule
\multirow{5}{*}{Places365}
& DNN $\ell_2$       & $0.9744$ & $0.9103$ \\
& DNN $\ell_\infty$  & $0.9341$ & $0.7912$ \\
& VLM $\ell_\infty$  & $0.7619$ & $0.7619$ \\
& LVLM $\ell_\infty$ & $0.6000$ & $0.6000$ \\
& Average            & $0.8176$ & $0.7659$ \\
\bottomrule
\end{tabular}
\end{table}

\section{Discussion}

\subsection{Potential Application Scenarios}
\label{sec:applications}

The evaluated setting suits workflows with gradient access but limited task data, metadata, or evaluation budgets.
\textbf{Third-party model auditing.} SGA uses few public probes for initial assessment and prioritizes models for comprehensive attacks.
\textbf{Private benchmark and competition pre-screening.} SGA screens submissions without exposing private test data or repeatedly running costly attacks, helping allocate budgets before standardized testing.
\textbf{Large-scale model development and selection.} Previously evaluated checkpoints serve as ordered anchors for Calibrated SGA to prioritize new candidates for full attack-based evaluation.
In all scenarios, SGA provides scalable first-stage assessment complementary to standardized attacks.

\subsection{Reference Robustness Orders}

Adversarial robustness has no protocol-independent ground truth, and attack-based orders depend on evaluation protocols. Although SGA is stable across closely related references (\cref{sec:reference-sensitivity}), \texttt{ViT-L-14\_TeCoA4} shows persistent disagreement in the VLM pool: PGD ranks it \#1, while Uncalibrated SGA ranks it \#5 across all probe sources. In \cref{tab:reference-outliers}, hypothetically moving only this model to \#5 substantially improves PA without changing other relative orders. This discrepancy offers a case for examining robustness properties captured differently by attack-based evaluation and structural sensitivity.

\begin{table}[!t]
\centering
\tablefontsize
\caption{\textbf{Persistent disagreement for \texttt{ViT-L-14\_TeCoA4}.} PGD ranks it \#1, whereas Uncalibrated SGA most often ranks it \#5; the rank-5 reference moves only this model while preserving all other orders.}
\label{tab:reference-outliers}
\setlength{\tabcolsep}{2.9mm}
\begin{tabular}{lcccc}
\toprule
Probe
& SGA rank
& Disagreement
& PGD ref.
& Rank-5 ref. \\
\midrule
Test set
& 5 (88\%)
& 100\%
& 0.7386
& 0.9195 \\
CelebA-HQ
& 5 (84\%)
& 99\%
& 0.8010
& 0.9695 \\
Places365
& 5 (100\%)
& 100\%
& 0.7686
& 0.9590 \\
\bottomrule
\end{tabular}
\end{table}


\section{Conclusion}
\label{sec:conclusion}

This work identifies a measurable structural signature of adversarial robustness in objective-conditioned local sensitivity. SGA captures this signature through structural alignment in a common spatial-structure domain, achieving strong ranking agreement across classifiers, VLMs, and LVLMs without prescribed attacks. With one input gradient per model--probe pair, SGA provides a practical auditing signal using few unlabeled, task-independent probes, limited model knowledge, and no adversarial-example generation. These findings establish structural sensitivity as an informative and operationally useful dimension of robustness beyond an optimization artifact.

SGA offers a lightweight alternative to expensive, benchmark-specific evaluation, supporting checkpoint screening and independent auditing before comprehensive attacks or certification. Prior work has established theoretical connections between input-gradient alignment and adversarial robustness~\cite{etmann2019connection}, providing a foundation for our structural sensitivity hypothesis. We further establish finite-probe concentration of SGA rankings and sufficient conditions for preserving their order across probe distributions. Nevertheless, extending these connections to SGA's structural alignment across architectures and threat models remains an open theoretical challenge. Addressing it could extend SGA beyond within-pool rankings toward a broader understanding of robustness through structural sensitivity.
\section*{Acknowledgments}

This work was partially supported by JSPS KAKENHI Grant JP24H00732, by JST CREST Grants JPMJCR20D3 and JPMJCR2562 including AIP challenge program, and by JST K Program Grant JPMJKP24C2 Japan.
\section*{Ethical Considerations}

SGA is a robustness-evaluation method, not an adversarial attack. It assesses models through their own structural sensitivity properties, without generating adversarial examples, exploiting vulnerabilities, or interacting with deployed systems. Our experiments use public datasets and model checkpoints without recruiting human participants. CelebA-HQ raises privacy and representation concerns; we use only unlabeled pixels, without identity, demographic, or attribute annotations or analysis, for non-commercial research under its terms. Although robustness scores could be misused to identify weaker models, SGA neither produces attack inputs nor provides an exploitation mechanism. Its intended use is preliminary security auditing, and its scores are not security guarantees or substitutes for attack-based or certified evaluation.
{
    \bibliographystyle{IEEEtran}
    \bibliography{references}
}

%

\appendix

\subsection{Methodological Differences and Challenges of LVLM Adaptation}
\label{app:method-comparison}

CLEVER~\cite{weng2018evaluating}, Linearized Robustness~\cite{etmann2019connection}, GREAT~\cite{li2024great}, and SGA are attack-agnostic: none requires adversarial examples or specific attacks. They assess different properties: CLEVER and Linearized Robustness estimate boundary distances, GREAT bounds distribution-level robustness via output confidence, and SGA measures spatial sensitivity organization. These differences affect cost, threat-model requirements, and LVLM adaptation.

\subsubsection{Core Algorithmic Differences}

Linearized Robustness divides the predicted-versus-competing-class margin by its input-gradient norm to estimate boundary distance. The margin represents class separation and its gradient local decision geometry; the first-order estimate uses one gradient at the input.

CLEVER extends this margin-over-sensitivity approach to a neighborhood. It repeatedly samples nearby points and their margin-gradient norms, then uses extreme-value estimation to approximate the maximum local cross-Lipschitz constant. Its neighborhood-based boundary-distance estimate costs substantially more gradients. Both methods require defined class margins and perturbation norms despite being attack-agnostic.

GREAT uses no input gradients or local boundary distances. It averages positive correct-versus-strongest-competitor confidence margins over class-conditioned generated samples. Under its assumptions, this lower-bounds the distribution-average $\ell_2$ attack-proof level. Its attack-agnostic interpretation requires known target classes, finite class sets, task-aligned generated samples, and its native $\ell_2$ formulation.

SGA uses the model's own prediction as a pseudo-target to extract an input gradient, constructs a sensitivity map, and compares it with probe-image structure. Unlike CLEVER and Linearized Robustness, it requires no output margin or gradient-norm boundary-distance interpretation. It is attack- and threat-norm-agnostic, generating neither adversarial examples nor norm-specific radius estimates.

\subsubsection{Differences in LVLM Adaptation}

GREAT, Linearized Robustness, and CLEVER assume closed-set classifiers with finite class scores, explicit targets, strongest competitors, and pairwise boundaries. Open-ended LVLMs instead generate autoregressive sequences without finite comparable response scores, a unique strongest competitor, or differentiable boundaries between complete answers; changing an early token can alter later prefixes and the full decoding trajectory. Thus, these methods lack native LVLM definitions.

For fair comparison, all three adaptations share one conditioning answer and sequence loss per model--probe pair. The sequence $\widetilde{\mathbf{y}}_{i,n}$ (\cref{eq:lvlm_probe_target}) is the ground-truth answer for test probes or clean greedy-decoded answer for public probes; its teacher-forced mean negative log likelihood (NLL) defines $\mathcal{L}_{i,n}$ (\cref{eq:lvlm_loss}). Their shared surrogate margin is
\begin{equation}
    m^{\mathrm{seq}}_{i,n}
    =
    \max\!\left(
        10-\mathcal{L}_{i,n},
        0
    \right).
    \label{eq:lvlm-surrogate-margin}
\end{equation}
The reference level 10 fixes the VQA-task surrogate's zero point; it is neither inherited from the original methods nor tuned to target-model robustness. Sharing it avoids method-specific competitors, extra supervision, and separately tuned LVLM parameters. GREAT scales $m^{\mathrm{seq}}_{i,n}$; Linearized Robustness divides it by the $\ell_2$ norm of the input gradient of $\mathcal{L}_{i,n}$; CLEVER divides it by the estimated local cross-Lipschitz constant of $10-\mathcal{L}_{i,n}$. Each retains its computational mechanism with shared conditioning and surrogate definitions.

These best-effort, consistent adaptations cannot recover the original closed-set assumptions. GREAT measures teacher-forced support for the chosen answer, not a margin over all responses; Linearized Robustness measures local first-order surrogate stability, not complete-answer boundary distance; CLEVER measures neighborhood surrogate stability, not a certified radius of invariant free-form output. These differences reflect absent native class margins, not unequal implementations.

SGA uses the same conditioning answer and teacher-forced $\mathcal{L}_{i,n}$ to compute $g_{i,n}$ (\cref{eq:input_gradient}). Treating gradients as spatial sensitivity rather than boundary geometry requires no competing-response margin; replacing classification with sequence loss preserves alignment between probe structure and objective-conditioned input sensitivity.

\subsection{Proofs of Fixed-Configuration Properties}
\label{app:fixed_configuration_proofs}

\subsubsection{Dice Mismatch Identity}

For binary maps $A$ and $B$, their symmetric difference satisfies
\begin{equation}
    |A\triangle B|
    =
    |A|+|B|-2|A\cap B|.
    \label{eq:symmetric_difference_identity}
\end{equation}
Therefore,
\begin{align}
    1-\operatorname{Dice}(A,B)
    &=
    1-
    \frac{2|A\cap B|}
    {|A|+|B|+\varepsilon_{\mathrm{D}}}
    \nonumber\\
    &=
    \frac{
        |A|+|B|-2|A\cap B|
        +\varepsilon_{\mathrm{D}}
    }
    {|A|+|B|+\varepsilon_{\mathrm{D}}}
    \nonumber\\
    &=
    \frac{
        |A\triangle B|
        +\varepsilon_{\mathrm{D}}
    }
    {|A|+|B|+\varepsilon_{\mathrm{D}}},
\end{align}
which proves \cref{eq:dice_mismatch}. Because the stabilizer appears in both expressions, the identity holds exactly, including for empty supports.

\subsubsection{Pairwise SGA-Order Concentration}

Fix a configuration $c$ independently of the evaluation probes. Let
\begin{equation}
    \mathcal{D}_1^{(1)},
    \ldots,
    \mathcal{D}_1^{(K)}
    \overset{\mathrm{i.i.d.}}{\sim}
    \mathcal{P}
\end{equation}
denote the $K$ independent single-probe batches that form $\mathcal{D}_K$, and define
\begin{equation}
    Y_n
    =
    S_c(f_i;\mathcal{D}_1^{(n)})
    -
    S_c(f_j;\mathcal{D}_1^{(n)}).
    \label{eq:proof_pairwise_variables}
\end{equation}
The variables $Y_1,\ldots,Y_K$ are independent. By the boundedness condition in \cref{eq:bounded_sga_score},
\begin{equation}
    Y_n\in[-r_c,r_c]
    \qquad
    \mathcal{P}\text{-almost surely}.
    \label{eq:proof_pairwise_range}
\end{equation}
Moreover,
\begin{align}
    \mathbb{E}[Y_n]
    &=
    \mathbb{E}_{\mathcal{D}_1^{(n)}\sim\mathcal{P}}
    \left[
        S_c(f_i;\mathcal{D}_1^{(n)})
        -
        S_c(f_j;\mathcal{D}_1^{(n)})
    \right]
    \nonumber\\
    &=
    \Theta_c(f_i;\mathcal{P})
    -
    \Theta_c(f_j;\mathcal{P})
    \nonumber\\
    &=
    \Delta_{ij,c}^{\mathcal{P}}.
    \label{eq:proof_pairwise_expectation}
\end{align}
By the definition of the SGA score in \cref{eq:sga},
\begin{align}
    &S_c(f_i;\mathcal{D}_K)
    -
    S_c(f_j;\mathcal{D}_K)
    \nonumber\\
    &\hspace{12mm}=
    \frac{1}{K}
    \sum_{n=1}^{K}
    Y_n.
    \label{eq:proof_empirical_pairwise_gap}
\end{align}
If $\Delta_{ij,c}^{\mathcal{P}}>0$, an empirical reversal or tie therefore occurs when
\begin{equation}
    \frac{1}{K}
    \sum_{n=1}^{K}Y_n
    \leq 0.
    \label{eq:proof_pairwise_inversion}
\end{equation}
Equivalently,
\begin{equation}
    \frac{1}{K}
    \sum_{n=1}^{K}Y_n
    -
    \Delta_{ij,c}^{\mathcal{P}}
    \leq
    -\Delta_{ij,c}^{\mathcal{P}}.
    \label{eq:proof_pairwise_deviation}
\end{equation}
Applying the one-sided Hoeffding inequality to the independent variables $Y_n$, each with interval width $2r_c$, gives
\begin{align}
    &\Pr_{\mathcal{D}_K\sim\mathcal{P}^K}\!\left[
        S_c(f_i;\mathcal{D}_K)
        \leq
        S_c(f_j;\mathcal{D}_K)
    \right]
    \nonumber\\
    &\leq
    \exp\!\left(
        -\frac{
            2K^2
            \left(\Delta_{ij,c}^{\mathcal{P}}\right)^2
        }{
            K(2r_c)^2
        }
    \right)
    \nonumber\\
    &=
    \exp\!\left(
        -\frac{
            K\left(\Delta_{ij,c}^{\mathcal{P}}\right)^2
        }{
            2r_c^2
        }
    \right),
\end{align}
which proves \cref{eq:pairwise_sga_concentration}.

For completeness, suppose that the $M$ models have distinct population SGA scores, and let
\begin{equation}
    \mathcal{I}_c^{\mathcal{P}}
    =
    \left\{
        (i,j):
        \Delta_{ij,c}^{\mathcal{P}}>0
    \right\}
    \label{eq:population_ordered_pairs}
\end{equation}
contain all population-ordered model pairs. By the union bound,
\begin{align}
    &\Pr_{\mathcal{D}_K\sim\mathcal{P}^K}\!\left[
        \exists(i,j)\in\mathcal{I}_c^{\mathcal{P}}:
        S_c(f_i;\mathcal{D}_K)
        \leq
        S_c(f_j;\mathcal{D}_K)
    \right]
    \nonumber\\
    &\leq
    \sum_{(i,j)\in\mathcal{I}_c^{\mathcal{P}}}
    \exp\!\left(
        -\frac{
            K\left(\Delta_{ij,c}^{\mathcal{P}}\right)^2
        }{
            2r_c^2
        }
    \right).
    \label{eq:full_sga_order_union_bound}
\end{align}
If
\begin{equation}
    \Delta_{\min,c}^{\mathcal{P}}
    =
    \min_{(i,j)\in\mathcal{I}_c^{\mathcal{P}}}
    \Delta_{ij,c}^{\mathcal{P}},
    \label{eq:minimum_population_sga_gap}
\end{equation}
then
\begin{align}
    &\Pr_{\mathcal{D}_K\sim\mathcal{P}^K}\!\left[
        \exists(i,j)\in\mathcal{I}_c^{\mathcal{P}}:
        S_c(f_i;\mathcal{D}_K)
        \leq
        S_c(f_j;\mathcal{D}_K)
    \right]
    \nonumber\\
    &\leq
    \binom{M}{2}
    \exp\!\left(
        -\frac{
            K\left(\Delta_{\min,c}^{\mathcal{P}}\right)^2
        }{
            2r_c^2
        }
    \right).
    \label{eq:full_sga_order_min_gap_bound}
\end{align}
Thus, when the population scores are distinct, the probability of any violation of their complete order is upper-bounded by a quantity that decreases exponentially with $K$.

\subsubsection{Probe-Distribution Order Preservation}

For any $f_i,f_j\in\mathcal{M}$,
\begin{align}
    &\left|
        \Delta_{ij,c}^{\mathcal{P}}
        -
        \Delta_{ij,c}^{\mathcal{Q}}
    \right|
    \nonumber\\
    &=
    \Big|
        \Theta_c(f_i;\mathcal{P})
        -
        \Theta_c(f_j;\mathcal{P})
        -
        \Theta_c(f_i;\mathcal{Q})
        +
        \Theta_c(f_j;\mathcal{Q})
    \Big|
    \nonumber\\
    &\leq
    \left|
        \Theta_c(f_i;\mathcal{P})
        -
        \Theta_c(f_i;\mathcal{Q})
    \right|
    +
    \left|
        \Theta_c(f_j;\mathcal{P})
        -
        \Theta_c(f_j;\mathcal{Q})
    \right|
    \nonumber\\
    &\leq
    2\operatorname{Disc}_c
    (\mathcal{P},\mathcal{Q};\mathcal{M}),
\end{align}
which proves \cref{eq:sga_gap_transfer}. Equivalently,
\begin{equation}
    \Delta_{ij,c}^{\mathcal{Q}}
    \geq
    \Delta_{ij,c}^{\mathcal{P}}
    -
    2\operatorname{Disc}_c
    (\mathcal{P},\mathcal{Q};\mathcal{M}).
    \label{eq:distribution_gap_lower_bound}
\end{equation}
Hence, if
\begin{equation}
    \Delta_{ij,c}^{\mathcal{P}}
    >
    2\operatorname{Disc}_c
    (\mathcal{P},\mathcal{Q};\mathcal{M}),
\end{equation}
then $\Delta_{ij,c}^{\mathcal{Q}}>0$, and therefore
\begin{equation}
    \Theta_c(f_i;\mathcal{Q})
    >
    \Theta_c(f_j;\mathcal{Q}).
\end{equation}
Thus, the population SGA order of this pair is preserved under the distribution change from $\mathcal{P}$ to $\mathcal{Q}$ whenever the condition in \cref{eq:sga_order_preservation} holds.

\subsection{Experimental Details}
\label{app:experimental-details}

\subsubsection{Data Construction and Preprocessing}
\label{app:data-details}

Classification test probes use all 1,000 NIPS development images from ImageNet-1K~\cite{nips2017}; public sets contain 1,000 CelebA-HQ~\cite{Lee2020mask} and 1,000 Places365~\cite{zhou2017places} images each. Public images are unlabeled and each model's initial prediction supplies its pseudo-target. All classification inputs are resized to $224\times224$.

VQA test probes use the first 1,000 VQAv2 validation questions~\cite{antol2015vqa}, corresponding COCO images~\cite{lin2014microsoft}, and ground-truth answers. Question-level sampling permits repeated COCO images. For public VQA probes, the $n$-th CelebA-HQ or Places365 image pairs with the $n$-th selected VQAv2 question, regardless of its original COCO image; each LVLM's greedy-decoded answer becomes the pseudo-target. Thus, neither public images nor their image--question pairings follow the task distribution.

LVLM inputs preserve aspect ratio: bilinearly resize the shorter side to the target resolution, then center-crop to a square. Transfer holds COCO, CelebA-HQ, and Places365 at $336\times336$. To test batch-specific selection across resolutions, Uncalibrated, Calibrated, and Oracle use $336\times336$ COCO, $160\times160$ CelebA-HQ, and $256\times256$ Places365. Models and methods share probes within each batch; batches draw different random probes.

\subsubsection{Model Pools and Reference Orders}
\label{app:models}

Four threat-homogeneous pools contain 39 entries from 38 distinct systems: DNN $\ell_2$ (13), DNN $\ell_\infty$ (14), VLM $\ell_\infty$ (7), and LVLM $\ell_\infty$ (5). Only \texttt{Standard\_R50} occurs twice, across the DNN pools. That checkpoint and the \texttt{ViT-L-14\_CLIP} encoder are not adversarially trained; all others use adversarially trained visual models or encoders. All five LVLMs use distinct visual encoders with the same LLaVA-1.5-7B interface~\cite{liu2023visual}.

Each PGD reference starts from 1,000 test probes (Appendix~\ref{app:data-details}) and retains only examples correct for every pool model: 611, 639, 513, and 686, respectively, giving all models 100\% clean accuracy on their PGD subsets. Single-restart PGD-50 uses seed 0, with $\epsilon=2$ and step $0.1$ for DNN $\ell_2$, or $\epsilon=8/255$ and step $1/255$ for all three $\ell_\infty$ pools.

DNN/VLM predictions use the highest class logit; PGD robustness is the retained fraction whose adversarial prediction matches the true class. LVLM PGD maximizes ground-truth-answer teacher-forced NLL; clean and adversarial answers use normalized exact match to ground truth. Robust accuracies induce separate within-pool orders, with no cross-pool magnitude comparison. \cref{tab:model-reference} lists exact identifiers, architectures, and descending PGD-50 values.

For DNN $\ell_\infty$, we also use standard RobustBench AutoAttack~\cite{croce2020reliable,croce2021robustbench} at $\epsilon=4/255$. \cref{tab:autoattack-reference} gives clean and robust accuracies, ordered by the latter.

\begin{table}[!t]
\centering
\tiny
\caption{\textbf{Models in descending PGD-50 robust accuracy.}}
\label{tab:model-reference}
\setlength{\tabcolsep}{0.15mm}
\begin{tabular}{llll}
\toprule
\textbf{Pool} & \textbf{Model identifier} & \textbf{Architecture} & \textbf{PGD reference} \\
\midrule

\multirow{13}{*}{DNN $\ell_2$}
& \texttt{Salman2020Do\_50\_2\_L2\_eps1.0}~\cite{salman2020adversarially}
& WideResNet-50-2 & $0.7529$ \\
& \texttt{Salman2020Do\_R50\_L2\_eps1.0}~\cite{salman2020adversarially}
& ResNet-50 & $0.7119$ \\
& \texttt{Salman2020Do\_R18\_L2\_eps1.0}~\cite{salman2020adversarially}
& ResNet-18 & $0.6154$ \\
& \texttt{Salman2020Do\_50\_2\_L2\_eps0.5}~\cite{salman2020adversarially}
& WideResNet-50-2 & $0.5516$ \\
& \texttt{Salman2020Do\_R50\_L2\_eps0.5}~\cite{salman2020adversarially}
& ResNet-50 & $0.5466$ \\
& \texttt{Salman2020Do\_R18\_L2\_eps0.5}~\cite{salman2020adversarially}
& ResNet-18 & $0.4141$ \\
& \texttt{Salman2020Do\_50\_2\_L2\_eps0.25}~\cite{salman2020adversarially}
& WideResNet-50-2 & $0.2848$ \\
& \texttt{Salman2020Do\_R50\_L2\_eps0.25}~\cite{salman2020adversarially}
& ResNet-50 & $0.2488$ \\
& \texttt{Salman2020Do\_R18\_L2\_eps0.25}~\cite{salman2020adversarially}
& ResNet-18 & $0.2079$ \\
& \texttt{Salman2020Do\_50\_2\_L2\_eps0.1}~\cite{salman2020adversarially}
& WideResNet-50-2 & $0.0933$ \\
& \texttt{Salman2020Do\_R50\_L2\_eps0.1}~\cite{salman2020adversarially}
& ResNet-50 & $0.0704$ \\
& \texttt{Salman2020Do\_R18\_L2\_eps0.1}~\cite{salman2020adversarially}
& ResNet-18 & $0.0442$ \\
& \texttt{Standard\_R50}~\cite{croce2021robustbench}
& ResNet-50 & $0.0000$ \\

\midrule

\multirow{14}{*}{DNN $\ell_\infty$}
& \texttt{Amini2024MeanSparse\_Swin-L}~\cite{amini2024meansparse}
& MeanSparse Swin-L & $0.6620$ \\
& \texttt{Amini2024MeanSparse\_ConvNeXt-L}~\cite{amini2024meansparse}
& MeanSparse ConvNeXt-L & $0.5900$ \\
& \texttt{Liu2023Comprehensive\_Swin-B}~\cite{liu2025comprehensive}
& Swin-B & $0.5430$ \\
& \texttt{Liu2023Comprehensive\_ConvNeXt-B}~\cite{liu2025comprehensive}
& ConvNeXt-B & $0.5039$ \\
& \texttt{Singh2023Revisiting\_ViT-B-ConvStem}~\cite{singh2023revisiting}
& ViT-B + ConvStem & $0.4695$ \\
& \texttt{Singh2023Revisiting\_ConvNeXt-S-ConvStem}~\cite{singh2023revisiting}
& ConvNeXt-S + ConvStem & $0.4554$ \\
& \texttt{Singh2023Revisiting\_ViT-S-ConvStem}~\cite{singh2023revisiting}
& ViT-S + ConvStem & $0.4038$ \\
& \texttt{Debenedetti2022Light\_XCiT-L12}~\cite{debenedetti2023light}
& XCiT-L12 & $0.3380$ \\
& \texttt{Debenedetti2022Light\_XCiT-M12}~\cite{debenedetti2023light}
& XCiT-M12 & $0.2989$ \\
& \texttt{Salman2020Do\_50\_2}~\cite{salman2020adversarially}
& WideResNet-50-2 & $0.2801$ \\
& \texttt{Debenedetti2022Light\_XCiT-S12}~\cite{debenedetti2023light}
& XCiT-S12 & $0.2269$ \\
& \texttt{Mo2022When\_Swin-B}~\cite{mo2022adversarial}
& Swin-B & $0.1878$ \\
& \texttt{Engstrom2019Robustness}~\cite{engstrom2019robustness}
& ResNet-50 & $0.1862$ \\
& \texttt{Standard\_R50}~\cite{croce2021robustbench}
& ResNet-50 & $0.0000$ \\

\midrule

\multirow{7}{*}{VLM $\ell_\infty$}
& \texttt{ViT-L-14\_TeCoA4}~\cite{mao2023understanding}
& ViT-L/14 & $0.3041$ \\
& \texttt{ConvNeXt-B\_RCLIP-TeCoA}~\cite{croce2025adversarially}
& ConvNeXt-B & $0.2437$ \\
& \texttt{ViT-B-16\_RCLIP-TeCoA}~\cite{croce2025adversarially}
& ViT-B/16 & $0.1793$ \\
& \texttt{ConvNeXt-B\_RCLIP-FARE}~\cite{croce2025adversarially}
& ConvNeXt-B & $0.1481$ \\
& \texttt{ViT-B-16\_RCLIP-FARE}~\cite{croce2025adversarially}
& ViT-B/16 & $0.0838$ \\
& \texttt{ViT-L-14\_LEAF-CLIP}~\cite{abad2026robustness}
& ViT-L/14 & $0.0351$ \\
& \texttt{ViT-L-14\_CLIP}~\cite{radford2021learning}
& ViT-L/14 & $0.0000$ \\

\midrule

\multirow{5}{*}{LVLM $\ell_\infty$}
& \texttt{ViT-L-14\_FARE4}~\cite{schlarmann2024robust}
& ViT-L/14 + LLaVA-1.5-7B & $0.9111$ \\
& \texttt{ViT-L-14\_SimCLIP4}~\cite{hossain2026sim}
& ViT-L/14 + LLaVA-1.5-7B & $0.8950$ \\
& \texttt{ViT-L-14\_LEAF-CLIP}~\cite{abad2026robustness}
& ViT-L/14 + LLaVA-1.5-7B & $0.8732$ \\
& \texttt{ViT-L-14\_TeCoA4}~\cite{mao2023understanding}
& ViT-L/14 + LLaVA-1.5-7B & $0.8528$ \\
& \texttt{ViT-L-14\_CLIP}~\cite{radford2021learning}
& ViT-L/14 + LLaVA-1.5-7B & $0.5758$ \\

\bottomrule
\end{tabular}
\end{table}

\begin{table}[!t]
\centering
\tiny
\caption{\textbf{DNN $\ell_\infty$ models in descending AutoAttack robust accuracy.} Standard AutoAttack results are from RobustBench at $\epsilon=4/255$.}
\label{tab:autoattack-reference}
\setlength{\tabcolsep}{0.15mm}
\begin{tabular}{llcc}
\toprule
\textbf{Model identifier}
& \textbf{Architecture}
& \textbf{Clean accuracy}
& \textbf{Robust accuracy} \\
\midrule
\texttt{Amini2024MeanSparse\_Swin-L}
& MeanSparse Swin-L
& $0.7880$ & $0.5892$ \\

\texttt{Amini2024MeanSparse\_ConvNeXt-L}
& MeanSparse ConvNeXt-L
& $0.7792$ & $0.5822$ \\

\texttt{Liu2023Comprehensive\_Swin-B}
& Swin-B
& $0.7616$ & $0.5616$ \\

\texttt{Singh2023Revisiting\_ViT-B-ConvStem}
& ViT-B + ConvStem
& $0.7662$ & $0.5466$ \\

\texttt{Liu2023Comprehensive\_ConvNeXt-B}
& ConvNeXt-B
& $0.7676$ & $0.5454$ \\

\texttt{Singh2023Revisiting\_ConvNeXt-S-ConvStem}
& ConvNeXt-S + ConvStem
& $0.7482$ & $0.5284$ \\

\texttt{Singh2023Revisiting\_ViT-S-ConvStem}
& ViT-S + ConvStem
& $0.7256$ & $0.4808$ \\

\texttt{Debenedetti2022Light\_XCiT-L12}
& XCiT-L12
& $0.7376$ & $0.4760$ \\

\texttt{Debenedetti2022Light\_XCiT-M12}
& XCiT-M12
& $0.7404$ & $0.4524$ \\

\texttt{Debenedetti2022Light\_XCiT-S12}
& XCiT-S12
& $0.7234$ & $0.4178$ \\

\texttt{Mo2022When\_Swin-B}
& Swin-B
& $0.7466$ & $0.3830$ \\

\texttt{Salman2020Do\_50\_2}
& WideResNet-50-2
& $0.6846$ & $0.3814$ \\

\texttt{Engstrom2019Robustness}
& ResNet-50
& $0.6256$ & $0.2922$ \\

\texttt{Standard\_R50}
& ResNet-50
& $0.7652$ & $0.0000$ \\
\bottomrule
\end{tabular}
\end{table}

\subsubsection{Baseline Configurations and Adaptations}
\label{app:baseline-details}

\cref{tab:baseline-settings} lists baseline settings. FGSM and CLEVER require the pool's $\ell_2$ or $\ell_\infty$ perturbation norm; mismatches worsen ranking. GREAT and Linearized Robustness need no norm. FMR lacks a public implementation and is excluded.

Originally evaluated on test inputs, all baselines use each model's initial prediction or greedy answer as pseudo-target for unlabeled public probes, matching SGA. FGSM uses step size $\epsilon$ and scores retention of the true or pseudo-target prediction.

Originally closed-set methods, FGSM, GREAT, CLEVER, and Linearized Robustness use the same LVLM conditioning answer $\widetilde{\mathbf{y}}_{i,n}$ and teacher-forced mean NLL $\mathcal{L}_{i,n}$ (\cref{eq:lvlm_probe_target,eq:lvlm_loss}). FGSM maximizes $\mathcal{L}_{i,n}$ and evaluates generated answers by normalized exact match. GREAT, CLEVER, and Linearized Robustness use the shared best-effort surrogate margin $m^{\mathrm{seq}}_{i,n}$ (\cref{eq:lvlm-surrogate-margin}): GREAT scales it, CLEVER estimates the local cross-Lipschitz constant of $10-\mathcal{L}_{i,n}$ by its original sampling and extreme-value procedure, and Linearized Robustness divides it by the $\ell_2$ norm of the input gradient of $\mathcal{L}_{i,n}$. SGA uses $\mathcal{L}_{i,n}$ to extract objective-conditioned gradients, with unchanged sensitivity representation, structural extraction, and similarity computation across DNNs, VLMs, and LVLMs. Appendix~\ref{app:method-comparison} explains the implications.

\begin{table}[!t]
\centering
\tiny
\caption{\textbf{Baseline settings.}}
\label{tab:baseline-settings}
\setlength{\tabcolsep}{0.06mm}
\begin{tabular}{l|l|l|l}
\toprule
Method & Required knowledge & Principal settings & Reported model score \\
\midrule
FGSM~\cite{goodfellow2014explaining}
& Perturbation norm ($\ell_2$ or $\ell_\infty$)
& $\epsilon$ matched to PGD-50
& Robust accuracy \\
GREAT~\cite{li2024great}
& None
& Shared default configuration
& GREAT score \\
CLEVER~\cite{weng2018evaluating}
& Perturbation norm ($\ell_2$ or $\ell_\infty$)
& \texttt{nsamp}$=50$; \texttt{niters}$=10$
& CLEVER score \\
Linearized Robustness~\cite{etmann2019connection}
& None
& \texttt{competitor}$=\text{top-2}$
& Linearized robustness score \\
\bottomrule
\end{tabular}
\end{table}

\subsubsection{SGA Candidate Space and Protocol Instantiation}
\label{app:configuration-selection}

All protocols share $\mathcal{C}$: $16\times12\times8\times3=4{,}608$ configurations (\cref{tab:sga-candidate-space}). Mutually exclusive smoothing options are $d=1$ (none), $d\in\{2,\ldots,7\}$ (bilinear down-/upsampling), or five Gaussian choices. Dice+SSIM averages both metrics. Candidates vary only structural post-processing and similarity; all reuse cached gradients without additional inference or backpropagation.

\begin{table}[!t]
\centering
\tablefontsize
\caption{\textbf{SGA configuration candidate space.} The Cartesian product of the four rows contains 4,608 configurations.}
\label{tab:sga-candidate-space}
\setlength{\tabcolsep}{3.4mm}
\begin{tabular}{ll}
\toprule
Component & Candidate values \\
\midrule
Gradient retention
& \shortstack[l]{$|g|\geq\{10^{-4},5\!\times\!10^{-5},10^{-5},5\!\times\!10^{-6},10^{-6}\}$;\\
top $\{50,55,\ldots,95\}$\%; or all} \\
\midrule
Smoothing
& \shortstack[l]{scale $d\in\{1,\ldots,7\}$; or\\Gaussian $\sigma\in\{0.3,0.4,\ldots,0.7\}$} \\
\midrule
Edge percentile
& $\{60,65,\ldots,95\}$ \\
\midrule
Similarity
& SSIM, Dice, or Dice+SSIM \\
\bottomrule
\end{tabular}
\end{table}

All selection follows \cref{eq:selection_cascade}. When multiple $K=10$ development batches are used, each criterion is averaged across batches before the next cascade layer, selecting one configuration per development condition rather than per-batch oracle configurations.

\emph{Transfer}
Two presets are selected on classification test probes and frozen before target evaluation. For the three $\ell_\infty$ target pools, development uses all $K=10$ test batches of DNN $\ell_2$ excluding \texttt{Standard\_R50}; its preset is edge percentile 95, top-90\% gradient retention, $d=2$, SSIM. DNN $\ell_2$ targets instead use DNN $\ell_\infty$ excluding \texttt{Standard\_R50} with the same batching; their preset is edge percentile 85, $|g|\geq10^{-5}$, Gaussian $\sigma=0.6$, Dice+SSIM. We exclude \texttt{Standard\_R50} to avoid overlap with either target pool. Each preset uses the full cascade (\cref{eq:selection_cascade}) across development batches and transfers unchanged to all models and three probe sources in its target pools. No target gradient, score, or reference order enters selection.

\emph{Uncalibrated}
For each pool and batch, \cref{eq:uncalibrated_selection} selects one shared configuration by score standard deviation, then minimum adjacent gap, then seeded random tie-breaking, without reference robustness.

\emph{Calibrated}
Except in the anchor-size ablation (\cref{sec:ablation_anchor}), each batch has $H=M-1$ anchors sampled without replacement with batch-index seed $b$ (seed 0 plus batch index), leaving one random target. We select via anchor scores and known order (\cref{eq:calibrated_selection}) and apply the result to the target. PA includes target--anchor and target--target pairs but excludes selection-used anchor--anchor pairs, also when $H<M-1$.

\emph{Oracle}
For each pool and batch, \cref{eq:oracle_selection} uses the full pool and reference order to choose one shared configuration independently per batch. It is only an empirical upper bound over evaluated probes and the finite candidate space.

\subsubsection{Batching, Reporting, and Runtime}
\label{app:runtime-details}

For $K\in\{1,2,5,10\}$, consecutive non-overlapping batches from the stored 1,000-example order yield 1,000, 500, 200, and 100 batches. All methods and models share batches within each pool/source; test-set, CelebA-HQ, and Places365 results remain separate. We report mean batch PA $\pm$ standard deviation per method, pool, source, and budget. This measures batch-composition variation, not confidence intervals or training randomness, since checkpoints are fixed.

Runtime uses one 94\,GB NVIDIA H100, measuring a full model pool and one $K=10$ batch. Wall-clock means follow one warm-up and three timed runs. Transfer and baselines include all steps through scoring; for Uncalibrated, Calibrated, and Oracle, ``$+$'' is extra selection/scoring time after shared gradients are cached.

\subsubsection{Ablation Protocols}
\label{app:ablation-protocols}

Unless specified, ablations use $K=10$, PGD-50 orders, the main evaluation's stored probe sequences and batches, and all three sources.

\emph{Spatial-structure destruction}
To isolate spatial organization from non-spatial gradient statistics, we reuse $K=10$ public-probe batches across four PGD-50 pools and two public sources. After caching $g_{i,n}$, a seeded pixel-location permutation moves each complete channel vector together:
\begin{equation}
    g^{\mathrm{perm}}_{i,n}[u,:]
    =
    g_{i,n}[\pi(u),:].
    \label{eq:spatial_permutation}
\end{equation}
This preserves per-pixel gradient-vector values and signs but destroys their spatial arrangement and probe correspondence. For each probe/seed, all pool models share the permutation. We use seeds 0--4, applying Uncalibrated SGA with the same candidate space to original and permuted gradients; permuted results are averaged across seeds.

\emph{Number of anchors}
For Calibrated SGA, $H$ varies from 2 to $M-1$ across four PGD-50 pools and three sources. Per batch and $H$, anchors are sampled without replacement using the main batch-indexed seeds; other models become targets. Selection uses only anchor scores and known order. PA includes all pairs with targets, excluding anchor--anchor pairs.

\emph{Architecture dependence}
This tests within-architecture ranking, not robustness differences between architectures. Models are grouped by architecture and homogeneous pool, without cross-pool PA. Zero-PGD-robust-accuracy checkpoints are excluded to prevent obvious endpoints from obscuring differences among robust models. Each subgroup/source selects one configuration from the full space via all $K=10$ batches and \cref{eq:selection_cascade}, fixed across batches. \cref{tab:architecture-ablation} gives subgroup compositions and sizes.

\emph{Configuration components}
To test component effects across pools/sources, one global base is selected using all $K=10$ batches across 12 PGD-50 conditions (four pools, three sources) and \cref{eq:selection_cascade}. It uses edge percentile 95, all gradient coordinates, $d=2$, and SSIM. We vary each component across its full range (\cref{tab:sga-candidate-space}) while fixing the others, reporting PA changes by pool/source.

\emph{Individual probes}
We examine visual properties of probe suitability separately on CelebA-HQ and Places365; test probes are excluded because pools lack a common image set. Each image receives independent $K=1$ Oracle SGA evaluation across four PGD-50 pools, yielding four PA values. Within each source, ranking uses mean PA, then maximum standard deviation, maximum minimum adjacent PA gap across those four values, and seeded random tie-breaking. The five best and five worst images form separate $K=5$ batches; each is evaluated with Uncalibrated SGA in each pool, jointly selecting one configuration for all five images in that pool.

\end{document}